\def\spose#1{\hbox to 0pt{#1\hss}}
\def\simlt{\mathrel{\spose{\lower 3pt\hbox{$\mathchar"218$}}
     \raise 2.0pt\hbox{$\mathchar"13C$}}}
\def\simgt{\mathrel{\spose{\lower 3pt\hbox{$\mathchar"218$}}
     \raise 2.0pt\hbox{$\mathchar"13E$}}}
\def\simpropto{\mathrel{\spose{\lower 3pt\hbox{$\mathchar"218$}}
     \raise 2.0pt\hbox{$\propto$}}}

\documentclass[useAMS,usenatbib]{mn2e}
\usepackage{amsmath,graphicx,bm,amsbsy}
\title{How well can we measure and understand foregrounds with $21\,\textrm{cm}$ experiments?}
\author[Adrian Liu, Max Tegmark]{Adrian Liu$^{1}$\thanks{E-mail:
acliu@mit.edu}, Max Tegmark$^{1}$\\
$^{1}$Dept. of Physics and MIT Kavli Institute, Massachusetts Institute of Technology, 77 Massachusetts Ave., Cambridge, MA 02139, USA}
\date{\today}
\begin{document}

\pagerange{\pageref{firstpage}--\pageref{lastpage}} \pubyear{2008}

\maketitle

\begin{abstract}
Before it becomes a sensitive probe of the Epoch of Reionization, the Dark Ages, and fundamental physics, $21\,\textrm{cm}$ tomography must successfully contend with the issue of foreground contamination.  Broadband foreground sources are expected to be roughly four orders of magnitude larger than any cosmological signals, so precise foreground models will be necessary.  Such foreground models often contain a large number of parameters, reflecting the complicated physics that governs foreground sources.  In this paper, we concentrate on spectral modeling (neglecting, for instance, bright point source removal from spatial maps) and show that $21\,\textrm{cm}$ tomography experiments will likely not be able to measure these parameters without large degeneracies, simply because the foreground spectra are so featureless and generic.  However, we show that this is also an advantage, because it means that the foregrounds can be characterized to a high degree of accuracy once a small number of parameters (likely three or four, depending on one's instrumental specifications) are measured.  This provides a simple understanding for why $21\,\textrm{cm}$ foreground subtraction schemes are able to remove most of the contaminants by suppressing just a small handful of simple spectral forms.  In addition, this suggests that the foreground \emph{modeling} process should be relatively simple and will likely not be an impediment to the foreground subtraction schemes that are necessary for a successful $21\,\textrm{cm}$ tomography experiment.
\end{abstract}

\begin{keywords}
Cosmology: Early Universe -- Radio Lines: General -- Techniques: Interferometric -- Methods: Data Analysis 
\end{keywords}

\maketitle

\section{Introduction}
In coming years, $21\,\textrm{cm}$ tomography has the potential to vastly expand our knowledge of both astrophysics and cosmology, with its unique ability to be a direct probe of the Dark Ages and the Epoch of Reionization \citep{Rees, Tozzi2, Tozzi, Iliev,furlanetto1,Loeb1,furlanetto2, Barkana1,Mack,Whitepaper1}, as well as its potential to provide constraints on cosmological parameters at an unprecedented level of accuracy \citep{Matt3,Santos2, juddjackiemiguel1,Yi,Whitepaper2,wyithe2008,ChangDE,miguelreview}.  First steps towards realizing this potential are being taken with the construction, testing, and initial science runs of low-frequency radio arrays such as PAPER \citep{PAPER}, LOFAR \citep{LOFARinstrument}, MWA \citep{MWAdesign}, CRT \citep{CRT}, GMRT \citep{GMRT}, and the Omniscope \citep{FFTT2}.  However, many obstacles still need to be overcome before cosmological measurements can be performed on the resulting data.  For instance, foreground contamination (both from within our galaxy and from extragalactic sources) is expected to be at least four orders of magnitude brighter than the expected cosmological signal \citep{angelica}, so any viable data analysis scheme must also contain a robust foreground subtraction algorithm.

Foreground subtraction is a problem that has been studied extensively in the literature, and many schemes have been proposed for tackling the issue.  Early ideas focused on using the angular structure of the foregrounds to separate them from the cosmological signal \citep{dimatteo1,dimatteo2,Oh,Santos,ZFH}, while most recent proposals focus on line-of-sight (\emph{i.e.} spectral) information \citep{xiaomin,nusserforegrounds,LOFAR,Harker,Judd08,paper2,paper1,LOFAR2,Petrovic}.    By making use of spectral information, these proposals take advantage of the extremely high spectral resolution available in all $21\,\textrm{cm}$ experiments, and indeed it was shown in \citet{paper4} that because of the nature of the foregrounds and the instrumental parameters, an optimal estimation of the power spectrum should involve a foreground subtraction scheme that operates primarily using frequency information.

Most line-of-sight proposals so far have been \emph{blind} schemes\footnote{See \citet{paper4} for an example that is \emph{not} blind.} in the sense that they do not require any prior foreground modeling.  All such proposals take advantage of the smooth nature of the foreground spectra, and separate out the rapidly fluctuating cosmological signal by (for instance) subtracting off a predetermined set of low-order polynomials \citep{Judd08,paper2,paper1} or by imposing a predetermined filter in Fourier space \citep{GMRT,Petrovic}.  The blind nature of these schemes may seem at first to be an advantage, because the low frequency ($\sim 50$ to $300\,\textrm{MHz}$) regime of radio foregrounds is as yet fairly unconstrained observationally and most models are based on extrapolations and interpolations from other frequencies, where the instruments are optimized for different science goals.  However, even if our foreground models are not entirely accurate initially, a non-blind scheme can always be performed iteratively until the models converge to the true measured foregrounds.  Moreover, blind schemes do not improve as one makes better and better measurements of the foregrounds, whereas non-blind schemes will continually improve as measurements place increasingly strong constraints on foreground models.  In this paper we therefore examine the foreground modeling process, and determine whether or not it will be possible to construct a foreground model that is good enough for foreground subtraction purposes if only a small number of independent parameters can be measured.  In the spirit of ``one scientist's noise is another scientist's signal", we also quantify the ability of $21\,\textrm{cm}$ experiments to place constraints on phenomenological foreground parameters.

The rest of this paper is organized as follows.  In Section \ref{fgmodel} we introduce our foreground model, and we show in Section \ref{fewparamsfine} that the foregrounds can be described to an extremely high precision using just three parameters.  This is good news for $21\,\textrm{cm}$ cosmology, for it implies that by measuring just a small handful of empirical parameters, it is possible to construct a foreground model that is sufficiently accurate for high precision foreground subtraction.  However, this also implies that the large number of (physically motivated) parameters present in a typical foreground parameterization are redundant and highly degenerate with each other, so the prospects for using $21\,\textrm{cm}$ experiments to gain a detailed understanding of foreground physics are bleak.  We show this in Section \ref{physicalparamsbad}, and summarize our conclusions in Section \ref{conc}.
\section{Foreground and Noise Model}
\label{fgmodel}
For the purposes of this paper, we limit our analyses to foregrounds in the frequency range $100$ to $250\,\textrm{MHz}$, which roughly speaking covers the ``sweet spots" of most $21\,\textrm{cm}$ experiments that are designed to probe the Epoch of Reionization.  At these frequencies, there are three dominant sources of foreground contamination \citep{Shaver,xiaomin}:
\begin{enumerate}
\item Extragalactic point sources.
\item Synchrotron emission from our Galaxy.
\item Free-free emission from our Galaxy.
\end{enumerate}
Since foreground subtraction is best done using spectral information \citep{paper4}, we will ignore the spatial structure of these foregrounds for this paper, as well as any polarization structure\footnote{See \citet{angelica} for a discussion of the spatial structure of foregrounds, and \citet{Bernardipol} and \citet{Geilpolarization} for discussions of polarized foregrounds and their removal.}.  We now examine the spectra of each of these sources, and in particular compute their means and variances, which we define in the next section.  With some minor modifications, our models are essentially those of \citet{paper4}.
\subsection{Definition of the mean and the foreground covariance}
Consider a multifrequency sky map of brightness temperature $T$ from a typical $21\,\textrm{cm}$ tomography experiment.  The map $T(\mathbf{\hat{r}},\nu)$ is a function of the direction in the sky $\mathbf{\hat{r}}$ as well as the frequency $\nu$.  We define a random function $x(\nu)$ such that
\begin{equation}
\label{xnu}
x(\nu) \equiv T(\mathbf{\hat{r}},\nu)
\end{equation}
is the spectrum measured at a randomly chosen pixel centered at $\mathbf{\hat{r}}$.  Since the cosmological signal is expected to be dwarfed by the foregrounds, this spectrum will essentially be a spectrum of the foregrounds.

To quantify the statistical properties of this random function, we can calculate its mean and covariance.  In principle, doing so requires taking the ensemble averages.  In practice, we instead take averages of the pixels of the sky map.  The mean is thus given by
\begin{equation}
\label{defmean}
m(\nu) = \langle x(\nu) \rangle \equiv \frac{1}{\Omega} \int T(\mathbf{\hat{r}},\nu) d\Omega,
\end{equation}
where $\Omega$ is the total area covered by the sky map\footnote{Note that this sky map need not cover all $4\pi$ steradians of the full sky.  For instance, in an estimation of the power spectrum one may choose to use only the data from the cleanest parts of the sky.}.  The statistical covariance can be similarly defined in the usual way:
\begin{equation}
\label{statcovar}
C(\nu,\nu^\prime) \equiv \langle x(\nu)x(\nu^\prime) \rangle - m(\nu)m(\nu^\prime),
\end{equation}
where
\begin{equation}
\langle x(\nu) x(\nu^\prime) \rangle = \frac{1}{\Omega} \int T(\mathbf{\hat{r}},\nu) T(\mathbf{\hat{r}},\nu^\prime) d\Omega.
\end{equation}

\subsection{Extragalactic point sources}
Extragalactic point sources can be thought of as consisting of two populations.  The first consists of bright, isolated point sources that can be resolved by one's instrument.  Following previous foreground studies, we assume that these bright point sources have already been removed,\ prior to the main foreground cleaning step that attempts to subtract off the rest of the foregrounds.  Techniques such as forward modeling \citep{forwardmodel} and peeling have been explored for this purpose, and peeling simulations suggest that the bright sources can be removed down to $S_{max} \sim 10$ to $100\,\textrm{mJy}$ \citep{Bart}.  To be conservative we will use $S_{max}=100\,\textrm{mJy}$ in all calculations that follow.

Below $S_{max}$ is the second population of extragalactic point sources, consisting of a ``confused" continuum of unresolved point sources.  Along a given line-of-sight, we imagine the number of sources to be Poisson distributed with an average of $n\Omega_{pix}$ sources, where $n$ is the number of sources per steradian and $\Omega_{pix}$ is the pixel size.  We model the spectrum of each source as a power law with a random spectral index $\alpha$ drawn from a Gaussian distribution
\begin{equation}
p(\alpha) = \frac{1}{\sqrt{2 \pi \sigma_\alpha^2}} \exp \left[ - \frac{(\alpha-\alpha_0)^2}{2 \sigma_\alpha^2} \right],
\end{equation}
where $\alpha_0$ is the mean spectral index (with its numerical value to be determined later) and $\sigma_\alpha =0.5$ \citep{max1}.  If all of the unresolved point sources had the same flux $S_*$ at some fiducial frequency $\nu_*\equiv 150 \,\textrm{MHz}$, the mean intensity would be
\begin{equation}
\label{meanpssingleflux}
 m^{ps}(\nu)  =  \left(\frac{A_\nu}{\Omega_{pix}}\right) \left( n \Omega_{pix} S_*\right) \int \left( \frac{\nu}{\nu_*} \right)^{-\alpha} p (\alpha) d \alpha,
\end{equation}
where the quantity $A_\nu/\Omega_{pix}$ converts our expression from having flux units to temperature units, and is given by
\begin{equation}
\left(\frac{A_\nu}{\Omega_{pix}}\right) = \frac{\lambda^2}{2 k_B \Omega_{pix}} \approx 1.4\times 10^{-6} \left( \frac{\nu}{\nu_*} \right)^{-2} \left( \frac{\Omega_{pix}}{1\,\textrm{sr}} \right)^{-1}\,\textrm{mJy}^{-1}\,\textrm{K},
\end{equation}
where $\lambda$ is the wavelength, $k_B$ is Boltzmann's constant, and $\Omega_{pix}$ is the pixel solid angle.  Of course, in reality the sources do not all have the same flux.  To take this into account, we extrapolate a source count distribution from an empirical study done at higher flux levels:
\begin{equation}
\frac{dn}{dS_*} = B \left( \frac{S_*}{880\,\textrm{mJy}}\right)^{-\gamma},
\end{equation}
where following \citet{dimatteo1} we take $B=4.0\,\textrm{mJy}^{-1}\textrm{Sr}^{-1}$ and $\gamma = 1.75$ as our fiducial values\footnote{Despite the fact that the source count distribution was obtained by extrapolation, we expect that whatever the true distribution is, it should be well-approximated by a power law.  This is because the key quantity is the integral of the source count, which is dominated by the brightest sources of the population.  In that regime, we can linearize the distribution in log-log space, giving a power law in $S_*$.}.  Integrating over the distribution, Equation \ref{meanpssingleflux} becomes
\begin{eqnarray}
 m^{ps}(\nu) &=& (17.4x_{max}^{2-\gamma}\,\textrm{K}) \left( \frac{B}{4.0 \,\textrm{mJy}^{-1}\textrm{Sr}^{-1}} \right) \times \nonumber \\
& &  \left( \frac{2-\gamma}{0.25} \right)^{-1}  \left(\frac{\nu}{\nu_*} \right)^{-\alpha_{ps} + \frac{\sigma_\alpha^2}{2} \ln \left( \frac{\nu}{\nu_*}\right)},
\end{eqnarray}
where $x_{max} \equiv S_{max}/880\,\textrm{mJy}$, and $\alpha_{ps} \equiv \alpha_0 + 2$ takes on a fiducial value of $2.5$ to match measurements from the Cosmic Microwave Background \citep{max1}.  Note also that this implies $\alpha_0 \approx 0.5$, which is consistent with both \citet{Toffolatti} and \citet{Jackson}.

To compute the covariance of the distribution, we once again begin by considering a population of point sources of brightness $S_*$ whose number density is determined by Poisson statistics (thus giving a result proportional to $nS_*^2$).  We then integrate over the source count distribution to get
\begin{eqnarray}
\label{Cps}
C^{ps} (\nu,\nu^\prime) \! \!\! \!\! \! &=& \! \!\! \!\! \! \frac{A_\nu A_{\nu^\prime}}{\Omega_{pix}} \int_0^{S_{max}} \frac{dn}{dS_*} S_*^2 dS_* \left( \frac{\nu \nu^\prime}{\nu_*^2} \right)^{- \alpha_0 + \frac{\sigma_\alpha^2}{2} \ln \left( \frac{\nu \nu^\prime}{\nu_*^2} \right)} \qquad \nonumber \\
&=&\! \!\! \!\! \! (4274x_{max}^{3-\gamma}\,\textrm{K}^2) \left( \frac{\Omega_{pix}}{10^{-6}\,\textrm{Sr}} \right)^{-1} \left( \frac{B}{4.0 \,\textrm{mJy}^{-1}\textrm{Sr}^{-1}} \right)  \nonumber \\
& & \times \left( \frac{3-\gamma}{1.25} \right)^{-1} \left(\frac{\nu \nu^\prime }{\nu_*^2} \right)^{-\alpha_{ps} + \frac{\sigma_\alpha^2}{2} \ln \left( \frac{\nu \nu^\prime}{\nu_*^2}\right)}.
\end{eqnarray}
Note that no further subtraction of the mean term is necessary, since it was implicitly accomplished when we invoked Poisson statistics.
\subsection{Galactic Synchrotron Radiation}
For Galactic synchrotron radiation, we imagine the foreground spectrum in each pixel to be well fit by a power law with spectral index $\alpha$, but that the value of the spectral index may vary from pixel to pixel.  In a given pixel, the spectrum is thus
\begin{equation}
x(\nu)= A_{sync} \left( \frac{\nu}{\nu_*} \right)^{-\alpha},
\end{equation}
where $A_{sync}=335.4\,\textrm{K}$ \citep{xiaomin}.  Similar to the point sources, we assume that the indices in different pixels to be Gaussian distributed, only this time with a mean of $\alpha_{sync}=2.8$ and a standard deviation of $\Delta \alpha_{sync}=0.1$ \citep{xiaomin}.  Performing the same integral as for the point sources, we obtain
\begin{equation}
 m^{sync}(\nu) = A_{sync} \left( \frac{\nu}{\nu_*} \right)^{-\alpha_{sync} +\frac{ \Delta \alpha_{sync}^2}{2} \ln \left( \frac{\nu}{\nu_*} \right) }.
\end{equation}
Forming the foreground covariance using Equation \ref{statcovar}, we obtain
\begin{eqnarray}
C^{sync} (\nu,\nu^\prime) =\!\!\!\! \!\!\!\! &&  A_{sync}^2 \left( \frac{\nu \nu^\prime}{\nu_*^2} \right)^{- \alpha_{sync} + \frac{ \Delta \alpha_{sync}^2}{2} \ln \left( \frac{\nu \nu^\prime}{\nu_*^2} \right)} \nonumber \\
& &- m^{sync}(\nu)  m^{sync}(\nu^\prime).
\end{eqnarray}
\subsection{Free-free Emission}
Free-free emission can be modeled in much the same way as the synchrotron radiation:
\begin{eqnarray}
 m^{f\!f}(\nu) =\!\!\!\! \!\!\!\! && A_{f\!f} \left( \frac{\nu}{\nu_*} \right)^{-\alpha_{f\!f} +\frac{\Delta \alpha_{f\!f}^2}{2} \ln \left( \frac{\nu}{\nu_*} \right) } \\
C^{f\!f} (\nu,\nu^\prime) =\!\!\!\! \!\!\!\! && A_{f\!f}^2 \left( \frac{\nu \nu^\prime}{\nu_*^2} \right)^{- \alpha_{f\!f} + \frac{ \Delta \alpha_{f\!f}^2}{2} \ln \left( \frac{\nu \nu^\prime}{\nu_*^2} \right)} \nonumber \\
& &- m^{f\!f}(\nu)  m^{f\!f}(\nu^\prime).
\end{eqnarray}
but with $A_{f\!f} = 33.5\,\textrm{K}$, $\alpha_{f\!f} = 2.15$, and $\Delta \alpha_{f\!f}=0.01$ \citep{xiaomin}.
\subsection{Total Foreground Contribution}
To obtain total contribution to the mean signal, we simply sum the means of the various components:
\begin{equation}
 m(\nu) =  m^{ps} (\nu) +  m^{sync} ( \nu ) +  m^{f\!f} (\nu).
\end{equation}
For the total covariance we sum the individual covariances:
\begin{equation}
C(\nu,\nu^\prime) = C^{ps} (\nu,\nu^\prime) + C^{sync} ( \nu,\nu^\prime ) + C^{f\!f} (\nu,\nu^\prime).
\end{equation}
In total, then, our foreground model consists of 10 free parameters, which are listed in Table \ref{table1} along with their fiducial values.
\begin{table}
\caption{Free parameters in our foreground model and their fiducial values.}
\begin{tabular}{p{1.5cm} p{4.0cm} p{1.8cm}}
\hline
\textbf{Parameter} & \textbf{Description} & \textbf{Fiducial Value}  \\
\hline
B & Source count normalization & $4.0\,\textrm{mJy}^{-1}\,\textrm{Sr}^{-1}$ \\
$\gamma $ & Source count power-law index & 1.75 \\
$\alpha_{ps}$ & Point source spectral index & 2.5 \\
$\sigma_{\alpha}$ & Point source index spread & 0.5 \\
$A_{sync}$ & Synchrotron amplitude & $335.4\,\textrm{K}$ \\
$\alpha_{sync}$ & Synchrotron spectral index & 2.8 \\
$\Delta \alpha_{sync}$ & Synchrotron index coherence & 0.1\\
$A_{f\!f}$ & Free-free amplitude & $33.5\,\textrm{K}$ \\
$\alpha_{f\!f}$ & Free-free spectral index & 2.15 \\
$\Delta \alpha_{f\!f}$ & Free-free index coherence & 0.01\\
\hline
\label{table1}
\end{tabular}
\end{table}
\subsection{Noise model}
\label{noisemodel}
In general, the noise level in a given pixel of a sky map produced by a radio telescope/interferometer scales as
\begin{equation}
\sigma_{noise} \propto \frac{\lambda^2 T_{sys}}{A_e \sqrt{\Delta \nu \Delta t}},
\end{equation}
where $\Delta \nu$ is the channel width of a frequency bin, $\Delta t$ is the integration time, $A_e$ is the collecting area of an antenna element, $T_{sys}$ is the system temperature, and $\lambda$ is the wavelength.  To keep the discussion in this paper as general possible, we do not explicitly model the effective area and the system temperature, since both depend on one's specific instrument in complicated ways.  Instead, we take a minimalistic approach and simply anchor our noise to levels that are expected for current generation $21\,\textrm{cm}$ tomography experiments.  From \citet{Judd08}, we know that with $360\,\textrm{hours}$ of integration at a channel width of $40\,\textrm{KHz}$, the MWA has a single pixel noise level at $158\,\textrm{MHz}$ that is approximately $330\,\textrm{mK}$.  Scaling this to our fiducial values, we obtain
\begin{equation}
\sigma_{noise} \sim (39\,\textrm{mK}) \left( \frac{1\,\textrm{MHz}}{\Delta \nu}\right)^{\frac{1}{2}} \left( \frac{1000\,\textrm{hrs}}{\Delta t}\right)^{\frac{1}{2}}.
\end{equation}
In the numerical studies conducted in this paper, we will always be keeping $\Delta t$ fixed at $1000\,\textrm{hrs}$.  The frequency bin width $\Delta \nu$ will also be fixed at its fiducial value of $1\,\textrm{MHz}$, except when explicitly stated (for instance when we investigate the dependence of our results on $\Delta \nu$).

In the next section, we will be working almost exclusively in dimensionless units where the foreground power (quantified by the diagonal $\nu=\nu^\prime$ elements of the foreground covariance) is equal to unity across all frequencies (see Equation \ref{nondim1}).  In such units, the noise can be taken to be approximately white \emph{i.e.} frequency independent.  To see this, note that the effective area $A_e$ of an antenna scales as $\lambda^2$, which means that the quantity $\lambda^2 / A_e$ has no frequency dependence, and $\sigma_{noise}$ depends on frequency only because of $T_{sys}$.  Typical $21\,\textrm{cm}$ experiments are sky noise dominated \citep{miguelreview}, which means that $T_{sys}$ should be dominated by the foreground temperature, and therefore should have roughly the same frequency dependence as the foregrounds do.  However, this frequency dependence was precisely the dependence that our choice of units was designed to null out.  The noise thus becomes white to a good approximation, and is effectively a noise-to-signal ratio (henceforth denoted by $\kappa$).  Note that while we adopt this approximation for the rest of the paper, it is by no means crucial, and in general one should always use units where the foreground covariance is white, even if in such units the noise is chromatic.  For further discussion of this point, please see Section \ref{eigenproperties}.

\section{The Ease of Characterizing Foregrounds}
\label{fewparamsfine}
While the foreground parameters described in Table \ref{table1} are certainly conventional choices, they are not the most economical, in the sense that many of them are redundant.  That this is the case is not surprising, since so many different foreground sources can be accurately described by spectra that deviate only slightly from power laws.  In Section \ref{eigensetup} we will reparametrize our foreground model using a principal component analysis.  In Section \ref{measurements} we will determine the effective number of principal components that need to be measured to accurately describe the foreground spectra, and find the number to typically be three or four.
\subsection{Eigenforeground Modes}
\label{eigensetup}
Since a measured foreground spectrum will necessarily be discrete, we begin by discretizing the mean and the covariance of our foreground model from Section \ref{fgmodel}, so that
\begin{equation}
\mathbf{m}_\alpha \equiv m(\nu_\alpha)
\end{equation}
and
\begin{equation}
\mathbf{C}_{\alpha \beta} \equiv C(\nu_\alpha, \nu_\beta),
\end{equation}
where the indices run from $1$ to $N_c$, the total number of frequency channels in one's instrument.  We define a correlation matrix
\begin{equation}
\label{nondim1}
\mathbf{R}_{\alpha \beta} \equiv \frac{\mathbf{C}_{\alpha \beta}}{\sqrt{\mathbf{C}_{\alpha \alpha} \mathbf{C}_{\beta \beta}}},
\end{equation}
and work with it instead the covariance matrix.

We now perform a principal component analysis on the foregrounds\footnote{We perform the principal component analysis using the correlation matrix $\mathbf{R}$ instead of the covariance matrix $\mathbf{C}$ because the challenge with foreground modeling is to successfully describe the fine relative perturbations about the smooth predictable power law spectrum.  Since the relative fluctuations are quantified by $\mathbf{R}$, not $\mathbf{C}$, we should correspondingly use $\mathbf{R}$ for our principal component analysis.}.  That is, we rewrite the correlation matrix in a basis of ``eigenforeground" vectors\footnote{Throughout this paper, we will use Greek indices to denote different frequencies and lowercase Latin indices to denote different foreground components/modes.  In Section \ref{physicalparamsbad} we will use uppercase Latin indices to denote the different foreground parameters listed in Table \ref{table1}.}, where each eigenforeground vector $\mathbf{v}_n$ satisfies the eigenvalue equation
\begin{equation}
\label{eigenvaleqn}
\mathbf{R}\mathbf{v}_n = \lambda_i \mathbf{v}_n.
\end{equation}
Normalizing the eigenforeground vectors to unity and forming a matrix $\mathbf{V}$ where the columns of the matrix are the normalized eigenvectors, the correlation matrix can be expressed as
\begin{equation}
\label{eigenbasis}
\mathbf{R} = \mathbf{V} \mathbf{\Lambda} \mathbf{V^t} =\sum_{n=1}^{N_c} \lambda_n \mathbf{v}_n \mathbf{v}_n^t,
\end{equation}
where $\mathbf{\Lambda} \equiv \textrm{diag}\{\lambda_1,\lambda_2,\lambda_3,\dots \}$.  Note that since $\mathbf{R}$ is real and symmetric, $\mathbf{V^t} = \mathbf{V}^{-1}$, so $\mathbf{V}\mathbf{V^t} = \mathbf{V^t} \mathbf{V} = \mathbf{I}$.  In the last equality of Equation \ref{eigenbasis}, we see that each eigenvalue $\lambda_n$ measures the contribution of its corresponding eigenforeground $\mathbf{v}_n$ to the total foreground variance.

In Section \ref{measurements}, we will seek to describe measured foreground spectra in terms of eigenforeground components.  In other words, we wish to find the weight vector $\mathbf{a}$ (with component $a_k$ for the $k^{th}$ eigenforeground) in the equation
\begin{equation}
\label{measurementeqn}
\mathbf{y}\equiv\mathbf{x} + \mathbf{n}=\sum_{k=1}^{N_c} a_k \mathbf{v}_k + \mathbf{n} = \mathbf{V} \mathbf{a}+ \mathbf{n} ,
\end{equation}
where $\mathbf{y}$, $\mathbf{x}$, and $\mathbf{n}$ are all vectors of length equal to the number of channels $N_c$, containing the measured spectrum of foregrounds, the true foregrounds, and the noise respectively.  The vector $\mathbf{x}$, for instance, is simply a discretized and whitened version of $x(\nu)$ in Equation \ref{xnu}.  Once an estimate $\mathbf{\hat{a}}$ of $\mathbf{a}$ has been determined, one can multiply by $\mathbf{V}$ to obtain an estimator $\mathbf{\hat{x}}$ of the foreground spectrum.  We will find that by measuring a small number of parameters, we can characterize foreground spectra to a very high precision, thanks to the special properties of the eigenforegrounds, which we describe in Section \ref{eigenproperties}.
%
%
\subsection{Features of the Eigenforegrounds}
\label{eigenproperties}
\begin{figure}
\centering
\includegraphics[width=0.45\textwidth]{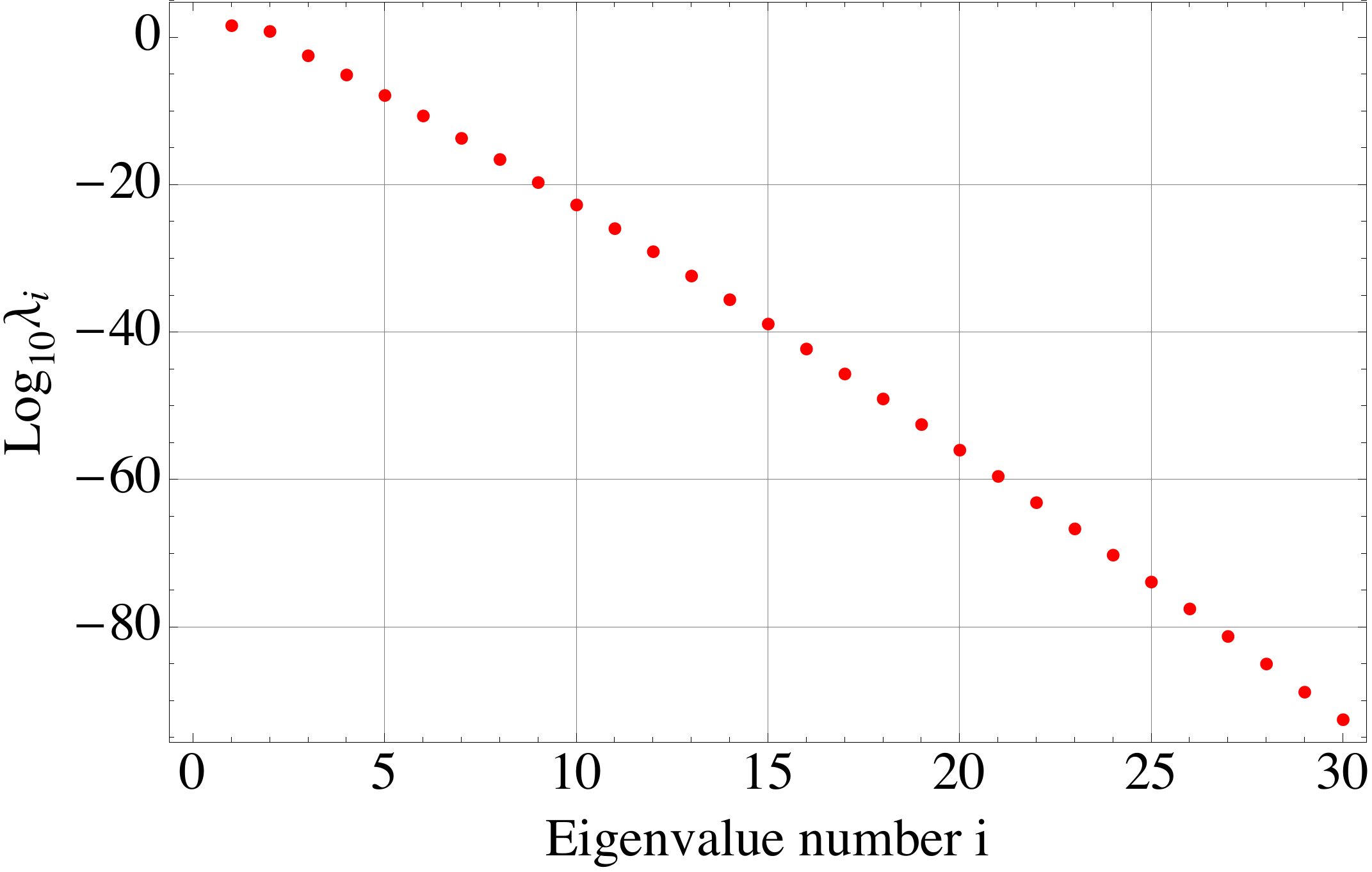}
\caption{Eigenvalues of $\mathbf{R}$ with no noise for an experiment with $50$ frequency channels, equally spaced from $100\,\textrm{MHz}$ to $200\,\textrm{MHz}$.}
\label{eigenvalueswnonoise}
\end{figure}
\begin{figure}
\centering
\includegraphics[width=0.45\textwidth]{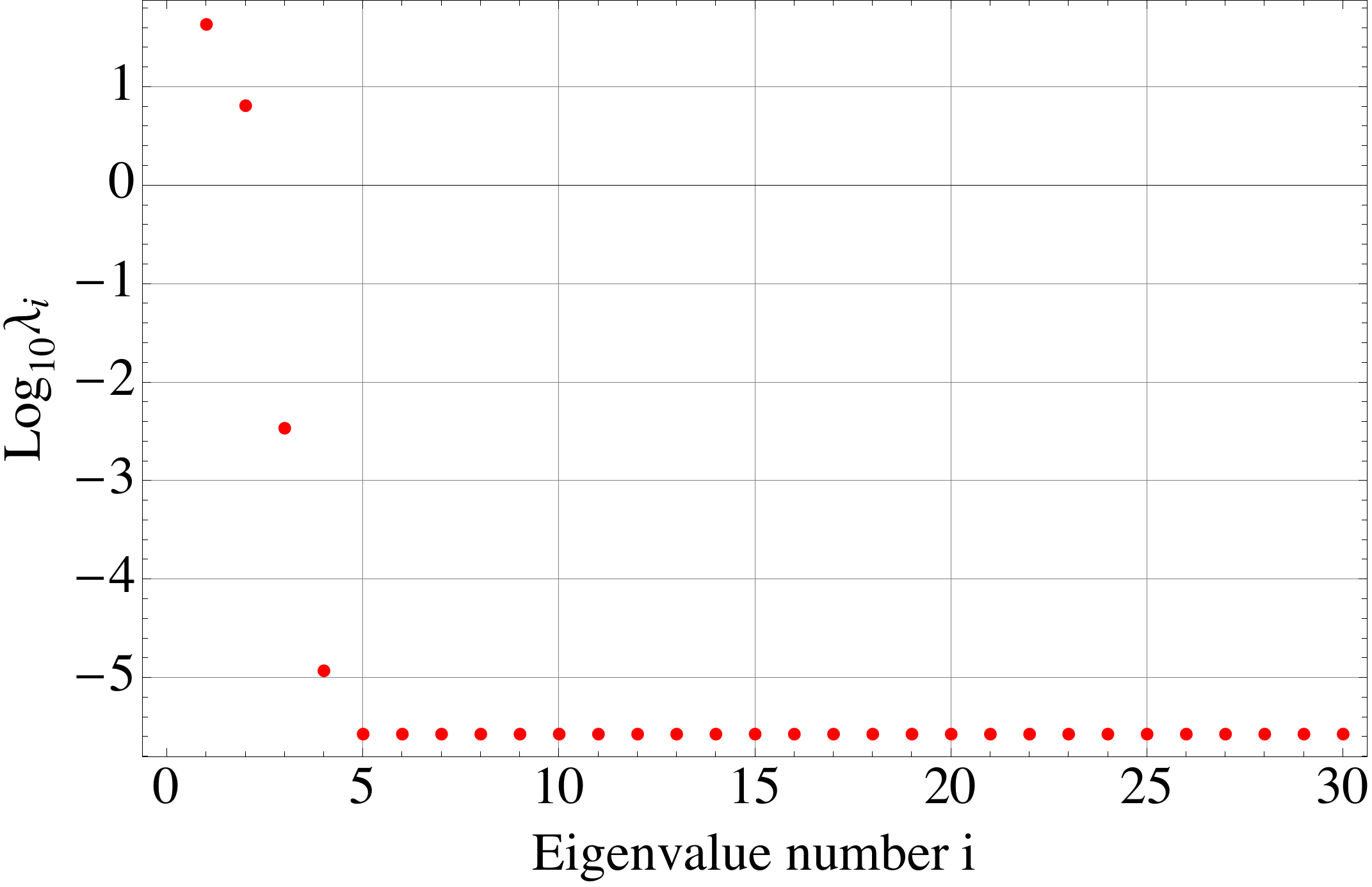}
\caption{Eigenvalues of $\mathbf{R}$ for an experiment with $50$ frequency channels, equally spaced from $100\,\textrm{MHz}$ to $200\,\textrm{MHz}$, and noise levels given by the fiducial model of Section \ref{noisemodel}.}
\label{eigenvalueswnoise}
\end{figure}
Since the foreground spectra do not contain any sharp features\footnote{Following \citet{miguelstatistical,nusserforegrounds}, we assume that narrowband contaminants such as terrestrial radio stations and radio recombination lines have been excised from the data prior to this point in the analysis.}, we expect their rather featureless frequency dependences to be well-described by a small number of eigenforegrounds, \emph{i.e.} we expect the eigenvalues $\lambda_n$ to be large only for the first few values of $n$.  Performing the analysis using the foreground model of Section \ref{fgmodel} for a noiseless experiment with $50$ equally spaced frequency channels going from $100\,\textrm{MHz}$ to $200\,\textrm{MHz}$, we see in Figure \ref{eigenvalueswnonoise} that the eigenvalues do fall off rapidly with mode number; indeed, the fall-off is exponential (a fact that we will explain later in this section), and in going from the first eigenvalue to the third there is a drop of more than three orders of magnitude.  This means that if we select as our foreground parameters the expansion coefficients $a_k$ of Equation \ref{measurementeqn}, we can account for almost all of the foreground signal by measuring just a few numbers.  In fact, in a realistic experiment it is impossible to measure more than the first few eigenvalues, as the foreground signal quickly becomes subdominant to the instrumental noise.  This can be seen in Figure \ref{eigenvalueswnoise}, where we once again show the eigenvalues, but this time including the noise model of Section \ref{noisemodel}.  After the fourth eigenforeground, we see that the eigenvalues hit a noise floor because by then the eigenmodes are essentially measuring the noise.  Note also that the large drop in magnitude of successive eigenvalues makes the qualitative results of this paper robust to our assumption that the noise is white in our non-dimensionalized units --- because the various eigenmodes contribute to the total foreground at such different levels, a slight chromaticity in the noise will not change the mode number at which the eigenmodes hit the noise floor in Figure \ref{eigenvalueswnoise}.  Violating our assumption will thus have very little effect on our results.

It should also be pointed out that in an analysis where the cosmological signal is included and is larger than the instrumental noise (as would be the case for a long integration time experiment), one would expect to hit a ``signal floor" rather than a noise floor.  The break in the eigenvalue spectrum in going from the exponential decay of the foreground dominated region to a flatter signal region can potentially be used as a diagnostic for separating foregrounds from signal.  To properly investigate the robustness of this method, however, requires detailed simulations of the signal.  Since the focus of this paper is foreground modeling (and not signal extraction), we defer such an investigation to future work.

\begin{figure}
\centering
\includegraphics[width=0.45\textwidth]{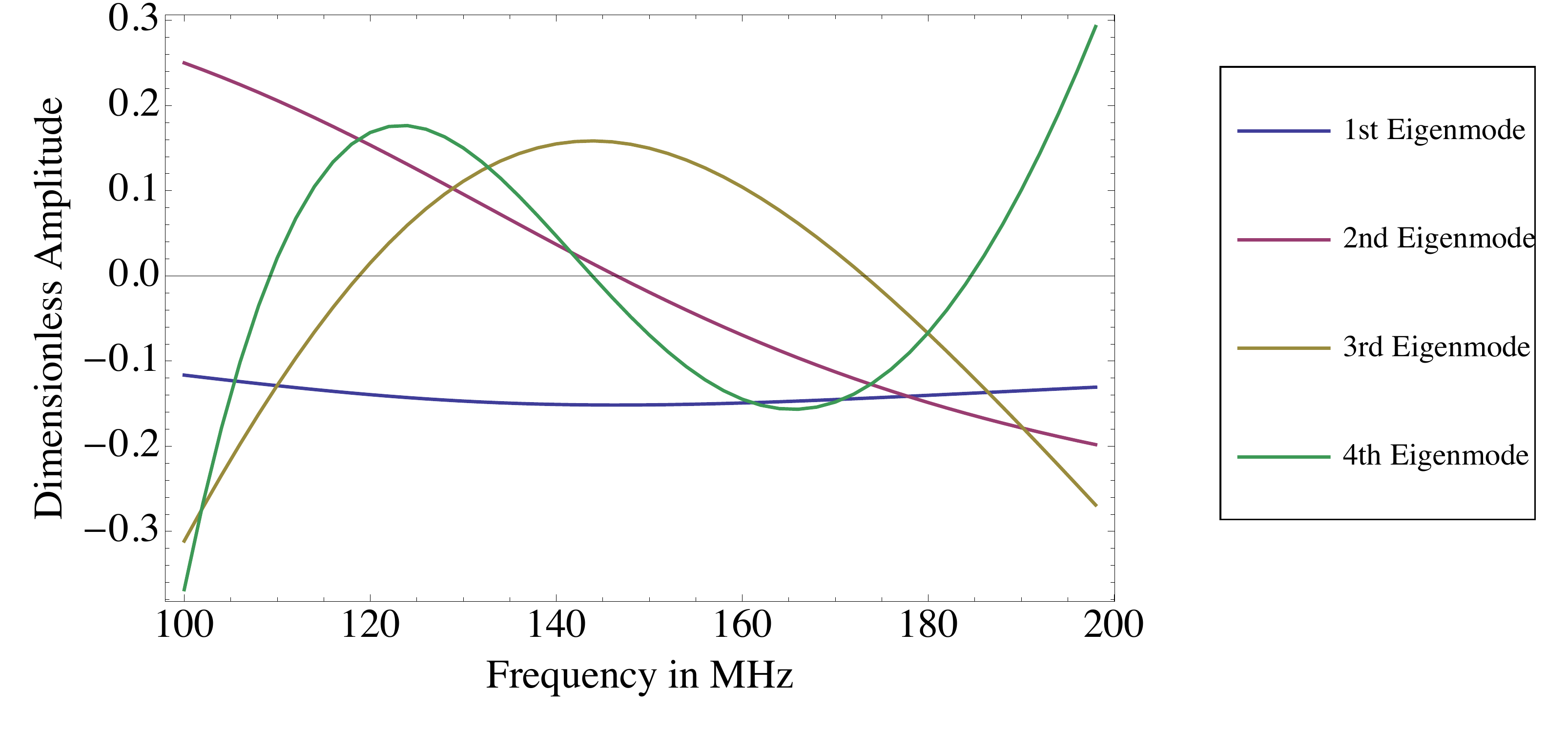}
\caption{First few eigenvectors of $\mathbf{R}$ (``eigenforegrounds") for an experiment spanning a frequency range from $100\,\textrm{MHz}$ to $200\,\textrm{MHz}$.}
\label{whitenedeigenforegrounds}
\end{figure}

\begin{figure}
\centering
\includegraphics[width=0.45\textwidth]{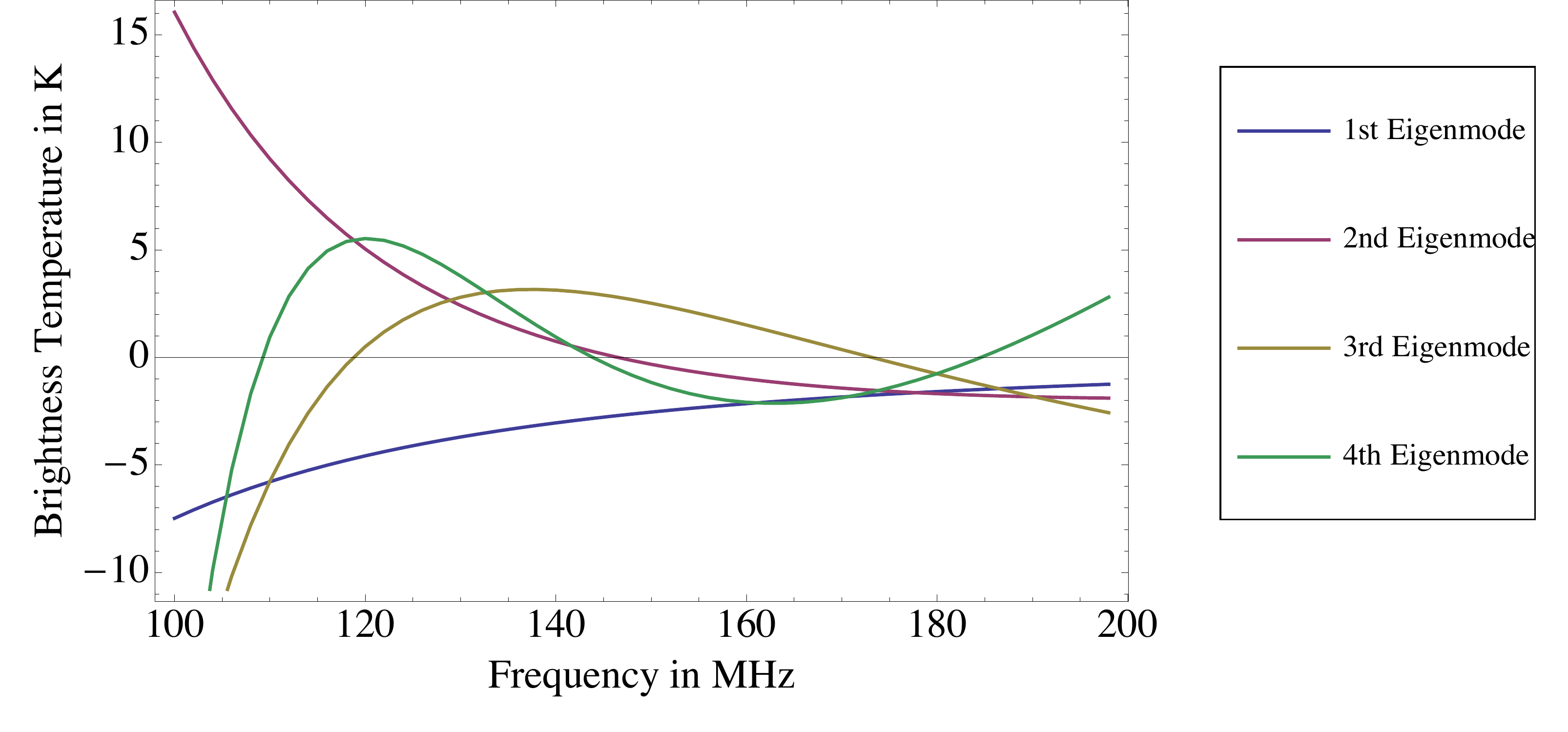}
\caption{First few eigenvectors of $\mathbf{R}$ (``eigenforegrounds") for an experiment spanning a frequency range from $100\,\textrm{MHz}$ to $200\,\textrm{MHz}$, but with the $\mathbf{C}_{\alpha \alpha}$ normalization factor restored so that the eigenforegrounds have units of temperature.}
\label{realunitseigenforegrounds}
\end{figure}

In Figure \ref{whitenedeigenforegrounds}, we show the first few foreground eigenvectors, and in Figure \ref{realunitseigenforegrounds} we restore the frequency-dependent normalization factors that were divided out in Equations \ref{nondim1}.  In both cases the eigenvectors are the ones defined by Equation \ref{eigenvaleqn} \emph{i.e.} for the noiseless case, and henceforth we will only be using these noiseless eigenvectors.  Even though we will include noise in our subsequent analysis, the use of the noiseless eigenvectors represents no loss of generality because the eigenvectors simply form a convenient set of basis vectors that span the space.  In addition, with our assumption of white noise (in the units defined by Equation \ref{nondim1}), the inclusion of noise alters only the eigenvalues, not the eigenvectors.

Several eigenforeground features are immediately apparent from the plots.  First, the $n^{th}$ eigenforeground is seen to have $n-1$ nodes.  This is due to the fact that correlation matrix $\mathbf{R}$ is Hermitian (since it is real and symmetric), so Equation \ref{eigenvaleqn} takes the mathematical form of a time-independent Schr\"{o}dinger equation for a one-dimensional system, and the node theorem of quantum mechanics applies.  As a consequence, each successive foreground eigenmode probes a more rapidly fluctuating component of the spectrum, which explains the rapid fall in eigenvalues seen in Figure \ref{eigenvalueswnonoise} --- since foregrounds are such smooth functions of frequency, the more rapidly oscillating eigenmodes are simply not required.


Another characteristic feature of Figure \ref{whitenedeigenforegrounds} is that the $n^{th}$ eigenforeground appears to look like a polynomial of order $(n-1)$, albeit with some slight deviations (the ``linear" 1st eigenmode, for instance, has a small curvature to it).  This approximate polynomial behavior explains the success of line-of-sight foreground subtraction schemes \citep{xiaomin,Judd08,paper2,paper1} that subtract low-order polynomials from foreground spectra --- by subtracting out low-order polynomials, one is subtracting the modes with the largest eigenvalues, and since the eigenvalues fall so quickly (indeed, exponentially) with mode number, the result is that most of the foregrounds are cleaned out by the process.  Together, Figures \ref{whitenedeigenforegrounds} and \ref{realunitseigenforegrounds} also explain why it was found in \citet{paper1} that polynomial foreground cleaning performs well only over narrow ($\sim 1\,\textrm{MHz}$) frequency ranges.  In Figure \ref{realunitseigenforegrounds}, it is seen that over large frequency ranges the eigenforegrounds do not behave like polynomials, and that the polynomial behavior is only evident after dividing out a power-law-like normalization.  In other words, the eigenforegrounds in Figure \ref{realunitseigenforegrounds} still have a rough power law dependence to their spectrum, and thus are not well-fit by polynomials.  Only if the frequency range is narrow will the power law dependence have a negligible effect, and so only in such a scenario would polynomial subtraction do well.  Fortunately, there is a simple solution to this problem --- if one wishes to perform polynomial subtraction over a wide frequency range, one can still do so successfully by first dividing out a fiducial spectrum (in the same way as we did in writing down Equation \ref{nondim1}) before performing the fits\footnote{Alternatively, one can fit foregrounds over a large frequency range by using polynomials in \emph{logarithmic} frequency space.  As noted in \citet{Judd08}, however, one may want to avoid doing this because of the interferometric nature of most $21\,\textrm{cm}$ experiments --- since interferometers are not sensitive to the mean emission, negative values will be measured in certain pixels, making it problematic to take the logarithm.}.  Put another way, we simply have to perform the fits in Figure \ref{whitenedeigenforegrounds} (where the modes \emph{always} look like polynomials) rather than in Figure \ref{realunitseigenforegrounds}.

\subsection{Understanding the Eigenforegrounds}
Despite the fact that it can be useful (in our whitened units) to think of the eigenforegrounds as polynomials of successively higher order, we caution that this interpretation is not exact.  For instance, the first eigenmode in Figure \ref{whitenedeigenforegrounds} is clearly not precisely constant, and contains a small quadratic correction.  Indeed, we will now show that it is more useful to think of the whitened eigenmodes as functions that would be sinusoids in $\nu$ were it not for an instrument's finite frequency range and spectral resolution.

Suppose that unresolved point sources were the sole contributor to our measured foregrounds.  Working in a continuous description (to avoid finite bandwidth and resolution), the analog of Equation \ref{nondim1} takes the form\footnote{Aside from the omission of the noise term here, what follows is precisely the toy model introduced in \citet{paper4}.}
\begin{equation}
R(\nu,\nu^\prime) = \frac{C(\nu,\nu^\prime)}{\sigma(\nu) \sigma(\nu^\prime)},
\end{equation}
where $C(\nu,\nu^\prime)$ is given by Equation \ref{Cps} and
\begin{equation}
\sigma(\nu) \equiv \left[ C(\nu,\nu) \right]^{\frac{1}{2}} = \left( \frac{\nu}{\nu_*} \right)^{-\alpha_{ps} + \sigma_\alpha^2 \ln ( \nu / \nu_* )}.
\end{equation}
The resulting correlation function $R(\nu, \nu^\prime)$ is given by
\begin{eqnarray}
\label{unnormedgaussian}
R(\nu, \nu^\prime) &=& \exp \left[ - \frac{\sigma_\alpha^2}{2} ( \ln \nu - \ln \nu^\prime)^2 \right] \nonumber \\
&\approx & \exp \left[ - \frac{(\nu - \nu^\prime )^2}{2 \nu_c^2}\right],
\end{eqnarray}
where we have Taylor expanded about $\nu_*$ to obtain an unnormalized Gaussian with a \emph{coherence length} of $\nu_c \equiv \nu_* / \sigma_\alpha \ln \nu_*$.  Despite the fact that this expression was derived by only considering point sources, we find that it is an excellent approximation even when the full foreground model is used, provided one uses a longer coherence length to reflect the smoother Galactic synchrotron and free-free components ($\sim64.8\,\textrm{MHz}$ gives a good fit to the foreground model of Section \ref{fgmodel}).

Since $R(\nu, \nu^\prime)$ now just depends on the difference $\nu-\nu^\prime$, the continuous analog of Equation \ref{eigenvaleqn} takes the form of a convolution:
\begin{equation}
\label{conv}
\int R(\nu - \nu^\prime) f_n(\nu^\prime) d\nu^\prime = \lambda_n f_n(\nu)
\end{equation}
where $f_n$ is the $n^{th}$ eigenforeground spectrum.  Because convolution kernels act multiplicatively in Fourier space, the $f_i$'s are a family of sinusoids.  Indeed, plugging the ansatz $f_n(\nu) = \sin (\gamma_n \nu + \phi)$ into Equation \ref{conv} yields
\begin{equation}
\int_{-\infty}^\infty  \exp \left[ - \frac{(\nu - \nu^\prime )^2}{2 \nu_c}\right] \sin (\gamma_n \nu^\prime +\phi) d\nu^\prime = \lambda_n \sin (\gamma_n \nu +\phi),
\end{equation}
where the eigenvalue is given by
\begin{equation}
\label{expeigen}
\lambda_n = \sqrt{2 \pi \nu_c^2} \exp\left(-2 \nu_c^2 \gamma_n^2 \right).
\end{equation}
The eigenforegrounds shown in Figure \ref{whitenedeigenforegrounds} should therefore be thought of as a series of sinusoid-like functions that deviate from perfect sinusoidal behavior only because of ``edge effects" arising from the finite frequency range of an experiment.  As expected from the exponential form of Equation \ref{expeigen}, the modes quickly decrease in importance with increasing wavenumber $\gamma_n$, and the more coherent the foregrounds (the larger $\nu_c$ is), the fewer eigenfunctions are needed to describe the foreground spectra to high accuracy.

One discrepancy between the eigenvalue behavior in our analytic treatment and the numerical results earlier on in the section is its dependence on wavenumber $\gamma_n$.  Analytically, Equation \ref{expeigen} predicts that $\gamma_n$ should appear quadratically in the exponent, whereas numerically Figure \ref{eigenvalueswnonoise} suggests something closer to a linear relationship (provided one imagines that the mode number is roughly proportional to $\gamma_n$, which should be a good approximation at high mode numbers).  We find from numerical experimentation that part of this discrepancy is due to the finite frequency range edge effects, and that as one goes from a very small range to a very large range, the dependence on $\gamma_n$ steepens from being linear in the exponential to being a power law in the exponential, with the index of the power law never exceeding $2$.  In any case, the role of Equation \ref{expeigen} is simply to provide an intuitive understanding of the origin of the exponential fall drop in eigenvalues, and in the frequency ranges applicable to $21\,\textrm{cm}$ tomography, we have verified numerically that a linear exponential fall-off fits the foreground model well.  In particular, we can parametrize the eigenvalues with two parameters $A$ and $B$, such that
\begin{equation}
\label{expdropoff}
\lambda_n \approx B \exp \left( - A n \right),
\end{equation}
and find that $A=6.44$ and $B=5.98\times10^5$ fits the foreground model well for an instrument with a frequency range of $100$ to $200\,\textrm{MHz}$ and frequency bins of $\Delta \nu = 2\,\textrm{MHz}$.  We caution, however, that these parameters vary with the frequency range, frequency bin width, and foreground model, and in general one must fit for $A$ and $B$ for each specific set of parameters.  This is what we do to generate the numerical results of the following subsection.  Generically, though, $A$ will always be somewhat larger than unity, so the foreground spectra will always be dominated by the first few eigenmodes.

\subsection{Eigenforeground Measurements}
\label{measurements}
So far, we have established that foreground spectra should be describable to a very high accuracy using only a small number of principal foreground components, with the importance of the $k^{th}$ foreground component quantified by the $a_k$ coefficient in $\mathbf{a}$, which was defined in Equation \ref{measurementeqn}.  In a practical measurement, however, the presence of noise means that one does not know the true value of $\mathbf{a}$, but instead must work with an estimator $\hat{\mathbf{a}}$ that is formed from the data.  There are many different ways to form this estimator, and one possibility would be to minimize the quantity
\begin{equation}
\chi^2 \equiv (\mathbf{y-V\hat{a}})^{\mathbf{t}} \mathbf{N}^{-1} (\mathbf{y-V\hat{a}}),
\end{equation}
where $\mathbf{N}$ is defined as $\langle \mathbf{n} \mathbf{n^t}\rangle $ using the noise vector $\mathbf{n}$ of Equation \ref{measurementeqn}.  This least-squares minimization (which is optimal if the noise is Gaussian) yields
\begin{equation}
\label{leastsq}
\mathbf{\widehat{a}}^{LS} = \mathbf{[V^t N^{-1} V]^{-1} V^t N^{-1} y =V^{-1} y = V^t y},
\end{equation}
since the number of eigenmodes (\emph{i.e.} the length of $\mathbf{\widehat{a}}^{LS} $) is equal to the number of frequency channels (\emph{i.e.} the length of $\mathbf{y}$), so that $\mathbf{V}$ is a square orthogonal matrix and $\mathbf{V}^t = \mathbf{V}^{-1}$.  Equation \ref{leastsq} states that even in the presence of noise, the least squares prescription calls for us to follow the same procedure as we would if there were no noise, namely, to take the dot product of both sides of Equation \ref{measurementeqn} with each eigenforeground vector.

Defining an error vector $\mathbf{\varepsilon} \equiv \delta \mathbf{a} = \mathbf{\hat{a} - a}$ as the difference between the true $\mathbf{a}$ and the estimator $\mathbf{\hat{a}}$, we could instead choose to minimize the diagonal quantities $\langle |\boldsymbol{\varepsilon}_i|^2 \rangle$.  This corresponds to so-called Wiener filtering \citep{mapmaking}, and the estimator is given by
\begin{equation}
\mathbf{\hat{a}}^{Wiener} = \mathbf{S V^t} \left[ \mathbf{VSV^t + N} \right]^{-1} \mathbf{y},
\end{equation}
where $\mathbf{S} \equiv \langle \mathbf{a}\mathbf{a^t} \rangle$ is the signal covariance matrix for the vector $\mathbf{a}$.

In Sections \ref{wiener} and \ref{TLS}, we examine Wiener filtering and least-squares minimization.  We will find that the estimated foreground spectrum $\mathbf{\hat{x}}=\mathbf{V \hat{a}}$ from Wiener filtering is expected to be closer to the true foreground spectrum, and that the least-squares method can be adapted to give similar results, although with errors larger than for Wiener filtering.

\subsubsection{Wiener Filtering}
\label{wiener}
With the Wiener filtering technique, our estimate of the foreground spectrum is given by
\begin{equation}
\mathbf{\hat{x}}^{Wiener} = \mathbf{V \hat{a}}^{Wiener} = \mathbf{V S V^t} \left[ \mathbf{VSV^t + N} \right]^{-1} \mathbf{y} \equiv \mathbf{W} \mathbf{y}
\end{equation}
Understanding Wiener filtering thus comes down to understanding the filter $\mathbf{W}$.  We first rewrite the expression slightly:
\begin{equation}
\mathbf{W} \equiv  \mathbf{V S V^t} \left[ \mathbf{VSV^t + N} \right]^{-1} = \mathbf{V S} \left[\mathbf{S} + \mathbf{V^t} \mathbf{N} \mathbf{V} \right]^{-1} \mathbf{V^t},
\end{equation}
where we have made liberal use of the fact that $\mathbf{V^t} = \mathbf{V}^{-1}$.  From Section \ref{noisemodel}, we know that $\mathbf{N}$ is proportional to the identity if we use whitened units.  In the notation of that section, we have $\mathbf{N} = \kappa^2 \mathbf{I}$, so
\begin{equation}
\mathbf{W} = \mathbf{V S} \left[\mathbf{S} + \kappa^2 \mathbf{I} \right]^{-1} \mathbf{V^t}.
\end{equation}
Now consider $\mathbf{S}$:
\begin{eqnarray}
\mathbf{S} &=&  \mathbf{V^t V S V^t V} =\mathbf{V^t} \langle \mathbf{(V a) (Va)^t} \rangle \mathbf{V} \nonumber \\
&=& \mathbf{V^t} \langle \mathbf{x x^t} \rangle \mathbf{V}= \mathbf{V^t R V}= \mathbf{\Lambda},
\end{eqnarray}
where $\mathbf{x}$ denotes the true whitened foreground spectrum, as it did in Equation \ref{measurementeqn}.  In the penultimate step we used the fact that $\langle \mathbf{x} \mathbf{x^t} \rangle$ is precisely our whitened foreground covariance, \emph{i.e.} $\mathbf{R}$, and in the last step we used Equation \ref{eigenbasis}.  Inserting this result into our Wiener filter gives
\begin{equation}
\label{Wienerfilter}
\mathbf{W}_{ij} =  \left( \mathbf{V \Lambda} \left[\mathbf{\Lambda} + \kappa^2 \mathbf{I} \right]^{-1} \mathbf{V^t} \right)_{ij}= \sum_{l,m}^{N_c} \mathbf{V}_{il} \left(\frac{\lambda_l }{\lambda_l + \kappa^2} \right) \delta_{lm} \mathbf{V^t}_{mj},
\end{equation}
where as before $N_c$ is the number of frequency channels (which is equal to the number of eigenforegrounds).  This in turn means that our estimator takes the form
\begin{eqnarray}
\mathbf{\hat{x}}^{Wiener}_i &=& \sum_{j}^{N_c} \mathbf{V}_{ij} \left(\frac{\lambda_j }{\lambda_j + \kappa^2} \right) (\mathbf{V^t y})_j \nonumber \\
&=&\sum_{j}^{N_c} \mathbf{V}_{ij} w_j \mathbf{\hat{a}}^{LS}_j=\sum_{j}^{N_c} w_j \mathbf{\hat{a}}^{LS}_j \mathbf{v}_j (\nu_i),
\end{eqnarray}
where $\mathbf{v}_j$ is the $j^{th}$ eigenforeground as defined in Equation \ref{eigenvaleqn} and $w_j \equiv \lambda_j/(\lambda_j + \kappa^2)$ can be thought of as ``Wiener weights" for the $j^{th}$ eigenforeground.  Since $w_j \approx 1$ for $\lambda_j \gg \kappa^2$ and $w_j \approx 0$ for $\lambda_j \ll \kappa^2$, this weighting factor preserves eigenmodes that have high signal to noise while suppressing those that have low signal to noise.  In words, the Wiener filtering procedure thus amounts to first performing a least-squares fit to obtain estimates for the eigenforeground coefficients, but then in the reconstruction of the foreground spectrum, downweighting modes that were only measured with low signal-to-noise.

\begin{figure}
\centering
\includegraphics[width=0.45\textwidth]{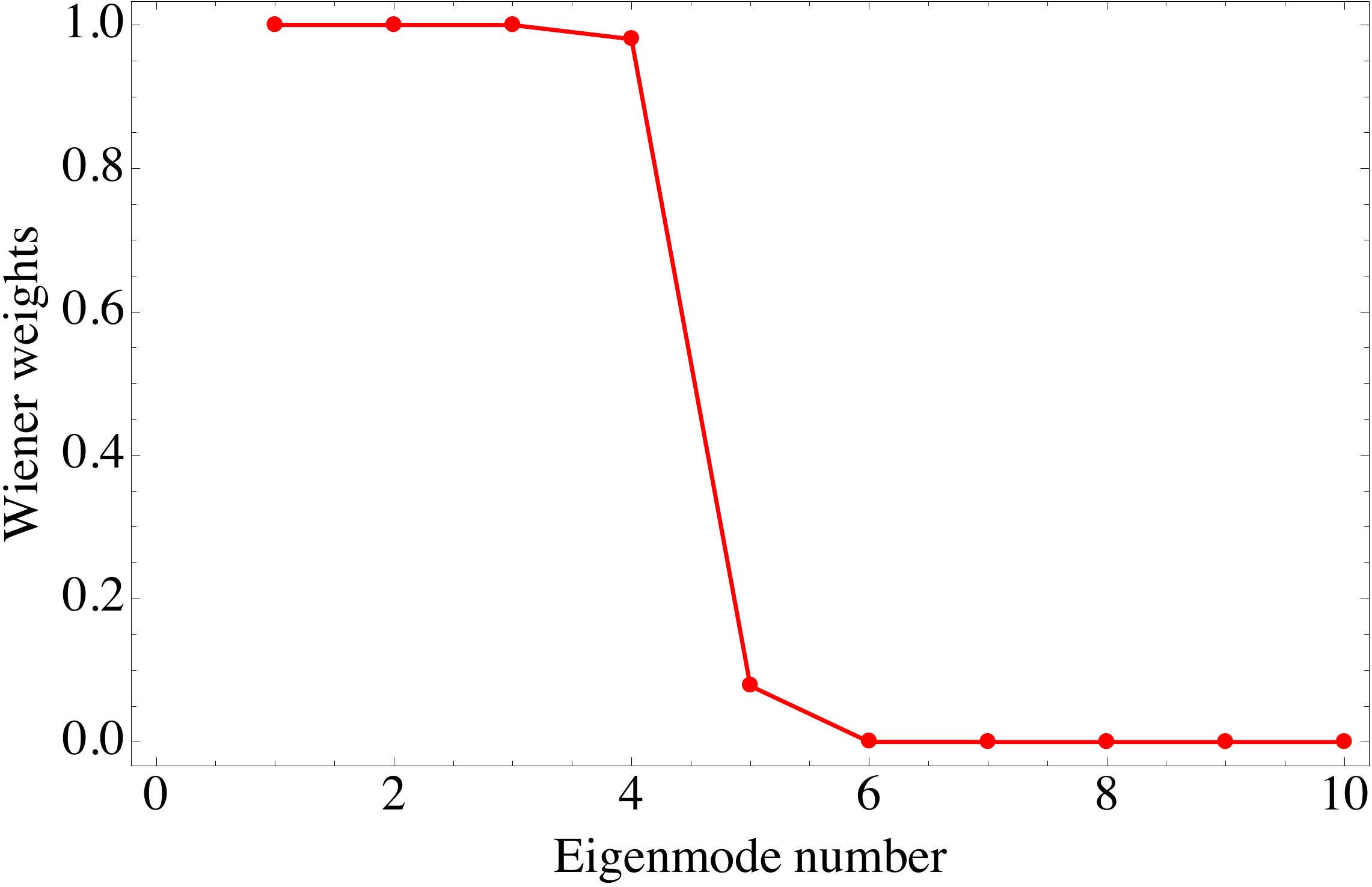}
\caption{First few Wiener weights (defined as $w=\lambda /(\lambda + \kappa^2)$, where $\lambda$ is the eigenvalue of a foreground mode and $\kappa$ is the whitened noise) for an experiment with $50$ frequency channels, equally spaced from $100\,\textrm{MHz}$ to $200\,\textrm{MHz}$, and noise levels given by the fiducial model of Section \ref{noisemodel}.}
\label{wienerweights}
\end{figure}

In Figure \ref{wienerweights} we show the Wiener weights for the first ten eigenmodes for an experiment with $50$ frequency channels, equally spaced from $100\,\textrm{MHz}$ to $200\,\textrm{MHz}$, and noise levels given by the fiducial model of Section \ref{noisemodel}.  The weights are seen to be small after the first few eigenmodes, suggesting that most of the information in our foreground spectrum estimate comes from a mere handful of parameters.  The other parameters (\emph{i.e.} eigenforeground coefficients) are too noisy to have much constraining power.  Since the Wiener weights tend to unity in the limit of large signal-to-noise, we can define an effective number of measurable parameters $n_{\textrm{eff}}$ by summing the Wiener weights:
\begin{equation}
\label{neff}
n_{\textrm{eff}} \equiv \sum_i^{N_c} w_i.
\end{equation}
For the fiducial scenario in Figure \ref{wienerweights}, the numerical value of $n_{\textrm{eff}}$ is 4.06.

In general, $n_{\textrm{eff}}$ depends on both the nature of the foregrounds and the noise level.  If the noise levels were high, the difficulty in measuring the higher (and therefore subdominant) eigenmodes would result in those measurements being noise dominated.  The Wiener filter would thus suppress such modes, driving $n_{\textrm{eff}}$ down.  In the opposite limit where the noise levels are low, one would expect $n_{\textrm{eff}}$ to be higher, but still to be relatively small.  This is because we know from Section \ref{eigenproperties} that the foreground spectra are relatively simple functions that are dominated by the first few eigenmodes, and thus even if the noise levels were low enough for the higher modes to be measurable, it is unnecessary to give them very much weight.  In general, $n_{\textrm{eff}}$ varies roughly logarithmically with increasing foreground-to-noise ratio since the foreground eigenvalues drop exponentially.  This nicely explains one of the results from \citet{paper4}.  There it was found that in performing foreground subtraction by deweighting foreground contaminated line-of-sight Fourier modes, the number of modes that needed to be deweighted increased only logarithmically with the foreground-to-noise ratio.  We now see that this is simply a consequence of Equation \ref{expdropoff}.

\begin{figure}
\centering
\includegraphics[width=0.45\textwidth]{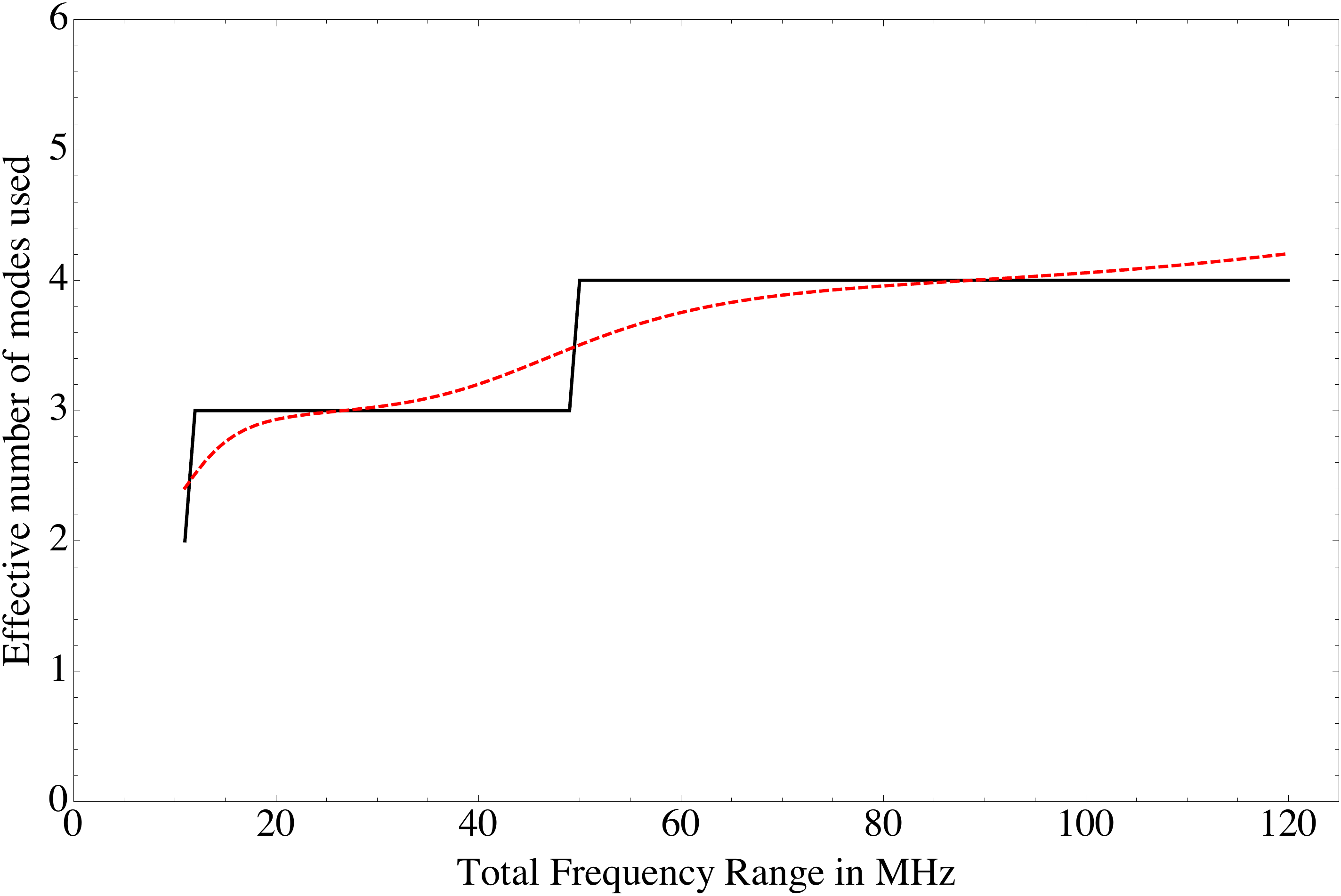}
\caption{Shown with the red/grey dashed curve, the effective number $n_{\textrm{eff}}$ of foreground parameters used (defined by Equation \ref{neff}) in the Wiener filtering method.  In solid black is the optimal $m$ (Equation \ref{optm} adjusted for the fact that $m$ must be an integer) for the truncated least-squares method.  In both cases the behavior is shown as a function of the total frequency range of an instrument, with channel width and integration time (and thus noise level) held constant at $1\,\textrm{MHz}$ and $1000\,\textrm{hrs}$, respectively.}
\label{optmvsNc}
\end{figure}

In Figure \ref{optmvsNc}, the red/grey dotted line shows $n_{\textrm{eff}}$ as a function of the total frequency range of an experiment.  In changing the frequency range, we keep the integration time and the channel width $\Delta \nu$ constant so that the noise level remains unchanged.  Any change in $n_{\textrm{eff}}$ therefore reflects the nature of the foregrounds, and what one sees in Figure \ref{optmvsNc} is that as the total frequency range increases, more and more modes are needed when estimating the spectrum.  This is because effects like the spread in the spectral indices are apparent only over large frequency ranges, so increasing the frequency range makes foregrounds more complicated to model.

To quantify the success of a foreground model generated from Wiener filtered measurements, we consider the quantity $\delta \mathbf{\hat{x}}^{w} \equiv \mathbf{\hat{x}}^{\textrm{Wiener}} - \mathbf{x}$:
\begin{eqnarray}
\label{wienererror}
\delta \mathbf{\hat{x}}^w_i &=& \sum_j^{N_c} \mathbf{V}_{ij} (w_j \mathbf{\hat{a}^{LS}}_j - \mathbf{a}_j)  \nonumber \\
&=& \sum_{j}^{N_c} \mathbf{V}_{ij}  (w_j -1) \mathbf{a}_j - \sum_{j,k}^{N_c} \mathbf{V}_{ij} w_j \mathbf{V}_{kj} \mathbf{n}_k \nonumber \\ 
&=& \sum_{j}^{N_c} (w_j -1) \mathbf{a}_j  \mathbf{v}_j- \sum_j^{N_c} \mathbf{W}_{ij} \mathbf{n}_j
\end{eqnarray}
where in the penultimate step we substituted $\mathbf{\hat{a}}^{LS} = \mathbf{a} + \mathbf{V^t}\mathbf{n}$ (which follows from Equations \ref{measurementeqn} and \ref{leastsq}), and in the ultimate step recognized that $\sum_{j} \mathbf{V}_{ij} w_j \mathbf{V}_{kj}$ is the Wiener filter $\mathbf{W}$ of Equation \ref{Wienerfilter}.  The covariance of this quantity is
\begin{eqnarray}
\label{wienercovar}
\langle \delta \mathbf{\hat{x}}^w \delta \mathbf{\hat{x}}_w^t \rangle \!\!\!\!&=&\!\!\!\! \sum_{i,j}^{N_c} \langle \mathbf{a}_i \mathbf{a}_j \rangle (w_i - 1) (w_j -1) \mathbf{v}_i \mathbf{v}_j^{\mathbf{t}} + \mathbf{W} \langle \mathbf{n} \mathbf{n^t} \rangle \mathbf{W^t} \nonumber \\
&=& \!\!\!\!\sum_{i}^{N_c} \lambda_i (w_i - 1)^2 \mathbf{v}_i \mathbf{v}_i^{\mathbf{t}} + \mathbf{W NW^t} \nonumber \\
&=& \!\!\!\!\sum_{i}^{N_c} \lambda_i (w_i - 1)^2 \mathbf{v}_i \mathbf{v}_i^{\mathbf{t}} + \kappa^2 \sum_i^{N_c} w_i^2 \mathbf{v}_i \mathbf{v}_i^{\mathbf{t}}
\end{eqnarray}
where in the penultimate step we used $\langle \mathbf{a}_i \mathbf{a}_j \rangle = \mathbf{S}_{ij} = \mathbf{\Lambda}_{ij} = \lambda_i \mathbf{\delta}_{ij}$, and in the last step used Equation \ref{Wienerfilter} as well as $\mathbf{N}= \kappa^2 \mathbf{I}$.  Note that there are no cross terms in going from Equation \ref{wienererror} to \ref{wienercovar} because such terms would be proportional to $\langle \sum_i a_i \mathbf{v}_i \mathbf{n^t} \rangle = \langle \mathbf{x} \mathbf{n^t} \rangle$, which is assumed to be zero because the foregrounds and the noise are uncorrelated.  The diagonal terms in Equation \ref{wienercovar} give the expected mean-square measurement error at a particular frequency. Averaging over frequency to obtain a figure-of-merit for the entire spectrum, our mean-square measurement error becomes
\begin{eqnarray}
\label{finalwienererror}
\varepsilon^2_{\textrm{Wiener}}  \!\!\!\!&\equiv& \overline{\langle \delta \mathbf{\hat{x}}^w \delta \mathbf{\hat{x}}_w^t \rangle} \nonumber \\
&=& \!\!\!\!\frac{1}{N_c}\sum_{\alpha}^{N_c} \sum_{i}^{N_c} \lambda_i (w_i - 1)^2 \mathbf{v}^2_i (\nu_\alpha)\nonumber\\
&& \qquad + \frac{\kappa^2}{N_c} \sum_{\alpha}^{N_c}  \sum_i^{N_c} w_i^2 \mathbf{v}^2_i (\nu_\alpha) \nonumber \\
&=& \frac{1}{N_c} \sum_i^{N_c} \lambda_i (w_i -1)^2+ \frac{\kappa^2}{N_c} \sum_i^{N_c} w_i^2,
\end{eqnarray}
where in the last step we permuted the sums and used the fact that the eigenforeground vectors are normalized.  From this expression, we can see that there are essentially two sources of error, each represented by one of the terms in Equation \ref{finalwienererror}.  The first term is only significant for the higher order eigenmodes, where $w_i \approx 0$.  With each element in the sum proportional to $\lambda_i$, this term quantifies the error incurred by heavily de-weighting the higher order modes (which are de-weighted so much that they are essentially omitted from the estimator).  Of course, we have no choice but to omit these modes, because their measurements are so noisy that there is essentially no foreground information contained in them.  The second term is significant only for the first few eigenforegrounds, which have $w_i \approx 1$.  Being proportional to $\kappa^2$, this term quantifies the error induced by instrumental noise in the measurement of the lower order modes (which are included in the estimator).

\begin{figure}
\centering
\includegraphics[width=0.45\textwidth]{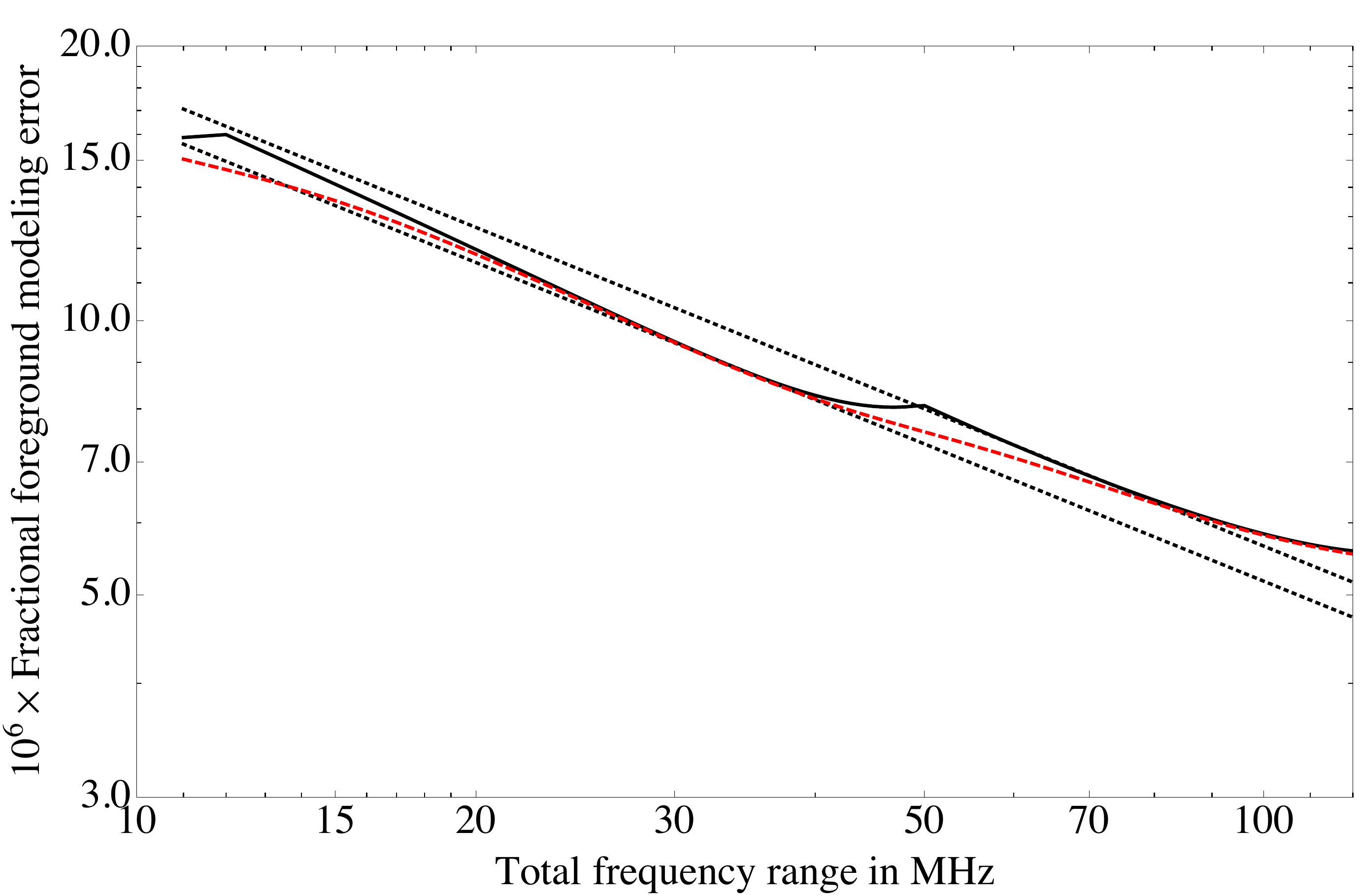}
\caption{Expected error on measured foregrounds divided by the root-mean-square (r.m.s.) foreground intensity to give a fractional foreground modeling error.  In dashed red/grey is the error for the Wiener filtering of Section \ref{wiener} (the square root of Equation \ref{finalwienererror} divided by the r.m.s.).  In solid black is the error for the truncated least-squares method of Section \ref{TLS} (Equation \ref{besterror} divided by the r.m.s. and adjusted for the fact that $m$ must be an integer).  In both cases we have plotted the errors on a log-log scale as functions of the total frequency range for an instrument with a fixed $1\,\textrm{MHz}$ channel width and $1000\,\textrm{hrs}$ of integration time, ensuring a constant noise level.  The black dotted lines are included for reference, and are proportional to $1/\sqrt{N_c \Delta \nu}$.}
\label{opterrorvsNc}
\end{figure}

In Figure \ref{opterrorvsNc}, we show (with the red/grey dashed line) the fractional error in an estimate of a foreground spectrum as a function of the total frequency range of an instrument.  The channel width $\Delta \nu$ and integration time $\Delta t$ are held constant, so there is no change in the noise level.  From the plot, we see that as the frequency range of the instrument increases, one is able to construct an increasingly accurate estimator of the foreground spectrum.  To understand why this is so, note that as the frequency range increases, there are two effects at play --- first, a larger frequency range means more complicated foregrounds, possibly increasing the errors in the fits; however, a larger frequency range with fixed channel width results in more data points to fit to, as well as a longer ``lever arm" with which to probe spectral features, thus reducing the errors.  What we see in Figure \ref{opterrorvsNc} is that the second outweighs the first.  This is unsurprising, since we saw from Figure \ref{optmvsNc} that as the frequency range is increased, the number of modes required to describe the foregrounds increases rather slowly, suggesting that the foreground spectra are only getting marginally more complicated.

\subsubsection{Truncated Least-Squares}
\label{TLS}
In a conventional least-squares fitting of a foreground spectrum, one forms an estimator for the spectrum by taking $\mathbf{\hat{a}^{LS}}$ and multiplying by $\mathbf{V}$.  In other words, one fits for all the eigenforeground coefficients (\emph{i.e.} the vector $\mathbf{a}$) and multiplies each coefficient by the spectrum of the corresponding eigenforeground.  This, however, is a suboptimal procedure for estimating the true spectrum because the higher order eigenmodes have very low signal-to-noise and measurements of them tend to be noise dominated.  Such modes should therefore be excluded (or heavily downweighted), which was what the Wiener filtering of the previous section accomplished.  In this section we explore a simpler method where we measure all eigenforeground coefficients, but include only the first $m$ eigenmodes in our estimate of the spectrum, truncating the eigenmode expansion so that the noisier measurements are excluded.

In this truncated least-squares scheme, our estimate of the spectrum for the $i^{th}$ frequency channel takes the form
\begin{equation}
\mathbf{\hat{x}}^{LS}_i=\sum_{j}^{m} \mathbf{\hat{a}}^{LS}_j \mathbf{v}_j (\nu_i) ,
\end{equation}
where $m$ is an integer to be determined later.  Written in this way, we see that we can reuse all the expressions derived in the previous section as long as we set $w_i = 1$ for $i \leqslant m$ and $w_i = 0$ for $i > m$.  Put another way, the truncated-least squares method approximates Wiener filtering by replacing the plot of Wiener weights shown in Figure \ref{wienerweights} with a step function.  Using Equation \ref{finalwienererror}, we thus see that the mean-square error for the truncated least-squares method is
\begin{eqnarray}
\label{finalLSerror}
\varepsilon^2_{\textrm{LS}}  &=& \frac{1}{N_c} \sum_{i=1}^{N_c} \lambda_i (w_i -1)^2+ \frac{\kappa^2}{N_c} \sum_{i=1}^{N_c} w_i^2 \nonumber \\
&=&  \frac{1}{N_c}\sum_{i=m+1}^{N_c} \lambda_i+ \frac{\kappa^2}{N_c}  \sum_{i=1}^{m} 1 \nonumber \\
&\approx& \frac{B}{N_c} \exp\left[-A(m+1) \right] + m \frac{\kappa^2}{N_c},
\end{eqnarray}
where in the last step we used the fact that the eigenvalues fall off in a steep exponential (Equation \ref{expdropoff}) that decays quickly enough that we can to a good approximation omit all but the first term in the sum.

From Equation \ref{finalLSerror}, we see that the truncated least-squares method gives large errors for extreme values of $m$, both large and small --- at large $m$, many modes are included, making the exponential term in the error small, but at the cost of large noise contamination from the second term; at low $m$, the noise term is small, but too few eigenforegrounds are included in the estimator of the spectrum, and the first term is large.  To obtain the best possible error from the truncated least-squares method, we must therefore choose an intermediate value of $m$ that minimizes $\varepsilon_{LS}^2$.  Differentiating with respect to $m$ and setting the result to zero allows one to solve for the optimal $m$:
\begin{equation}
\label{optm}
m_{opt} = \frac{1}{A} \ln \left( \frac{AB}{\kappa^2} \right) -1,
\end{equation}
and inserting this into our expression for the error gives a best error of
\begin{equation}
\label{besterror}
\varepsilon_{LS}^{best} \sim \frac{\kappa_{fid}}{\sqrt{N_c \Delta \nu\Delta t }} \left[ \frac{1}{A} \left[ \ln \left( \frac{AB  \sqrt{\Delta t  \Delta \nu}}{\kappa_{fid}^2} \right)  +1 \right] -1 \right]^{\frac{1}{2}},
\end{equation}
where we have made the substitution $\kappa \rightarrow \kappa_{fid}/\sqrt{\Delta t \Delta \nu}$ to emphasize the scalings with integration time $\Delta t$ and channel width $\Delta \nu$.  Note that Equation \ref{besterror} is only approximate, since $m$ must take on an integer value.

In Figure \ref{optmvsNc}, the solid black curve shows $m_{opt}$ as a function of $N_c \Delta \nu$, fitting numerically for $A$ and $B$ (Equation \ref{expdropoff}) for each $N_c \Delta \nu$.  The channel width $\Delta \nu$ and $\Delta t$ are held at $1\,\textrm{MHz}$ and $1000\,\textrm{hrs}$ respectively, as described in Section \ref{noisemodel}.  Since $\Delta \nu$ and $\Delta t$ are constant, the normalized noise level $\kappa$ is also constant, and the variation in $m_{opt}$ is due purely to changes in $A$ and $B$.  This variation is seen to be quite weak, and we see that the optimal number of components to solve for is always a rather small number, and that this number increases only slowly with the number of channels $N_c$ (or equivalently with the total frequency range $N_c \Delta \nu$, since the channel width $\Delta \nu$ is being held constant).  This is unsurprising, given that the first term in Equation \ref{finalLSerror} decays exponentially while the second term rises only linearly, forcing the optimal $m$ towards small $m$.  With any current-generation experiment, we find $m_{opt}$ to be no larger than $4$.  Also note that $m_{opt}$ behaves like a discretized version of $n_{\textrm{eff}}$ from the Wiener filtering, as is expected.

In Figure \ref{opterrorvsNc}, we show how the optimal $m$ values of Figure \ref{optmvsNc} translate into the error $\varepsilon_{total}^{best}$, again as a function of $N_c \Delta\nu$.  The error is divided by the root-mean-square value of the foregrounds over the entire spectrum to give a rough percentage error, and is plotted as the solid black line, while reference lines proportional to $1/\sqrt{N_c \Delta \nu}$ are shown in dotted black.  The error is seen to decrease mostly with the inverse square root of the total frequency range $N_c \Delta \nu$, as one would expect from the prefactor in Equation \ref{besterror}.  The deviation from this behavior is due to the fact that the second part of Equation \ref{besterror} depends on $A$ and $B$, which in turn vary with $N_c$.  This, however, is a rather weak effect, since the relevant parts of Equation \ref{besterror} look a lot like our expression for $m_{opt}$, which we know from Figure \ref{optmvsNc} rises slowly.  The result is that deviations from a $1/\sqrt{N_c \Delta \nu}$ scaling occur only when one is close to the transition in $m_{opt}$ (which remember, must be an integer).  Again, we see that that truncated least-squares method closely approximates the Wiener filtering method, with only slightly larger errors.

With both methods, we find that there is very little change in the error if one changes the channel width $\Delta \nu$ while correspondingly adjusting $N_c$ so that $N_c \Delta \nu$ (\emph{i.e.} the total frequency range of the instrument) is kept constant.  For the truncated least-squares method, such changes keep the prefactor in Equation \ref{besterror} fixed, and any variations in the error are due purely to the corrections in the second part of the equation.  Numerically, the lack of sharp spectral features in the foregrounds means that these corrections are found to be completely subdominant.  This implies that to an excellent approximation, the errors in our measured foregrounds are dependent only on the total frequency range of our instrument, and are independent of how we bin our data --- binning more coarsely results in a lower noise per frequency bin, but this is canceled out (to a very high precision) by the larger number of bins with which to perform our fits\footnote{This is provided one makes the (usually excellent) assumption that the number of frequency bins is much larger than the number of independent foreground modes, seen in this section and the previous one to be $3$ or $4$.}.  From this, we see that the decrease in errors seen in Figure \ref{opterrorvsNc} (where we increased the number of data points by increasing the total frequency range of the instrument) is due not to intrinsically better fits, but rather to a longer ``lever arm" over which to probe foreground characteristics.

\subsubsection{Which method to use?}
 In summary, we can see from Figure \ref{optmvsNc} that it is indeed the case that $21\,\textrm{cm}$ foregrounds can be accurately characterized using just a small number ($\sim3$ or $4$) of independent components.  From Figure \ref{opterrorvsNc}, we see that both Wiener filtering and the truncated least-squares method allow foregrounds to be estimated to an accuracy of roughly one part in $10^{-5}$ to $10^{-6}$.  This is fortunate since the cosmological signal is expected to be $\sim10^{-4}$ smaller than the foregrounds, so a failure to reach at least that level of precision would ruin any prospects of a cosmological measurement.
 
 With the Wiener filtering and the truncated least-squares giving such similar results, for most applications it should not matter which method is used.  However, if one requires the very best foreground model achievable, then Wiener filtering should be used, since it was derived by minimizing the error.  On the other hand, the least-squares method has the advantage of being simpler, and in addition is more immune to inaccuracies in foreground modeling.  This is because Wiener filtering explicitly involves the foreground covariance $\mathbf{S}$, and for the method to minimize the modeling error, it is important that the foreground-to-noise ratio be accurate.  On the other hand, the foregrounds only enter the least squares method via the choice of basis (\emph{i.e.} through $\mathbf{V}$), and since this basis spans the vector space, the role of the foreground model in least-squares fitting is simply that of a prior model.  Ultimately, though, the Wiener filtering method's need for an estimate of the foreground covariance is unlikely to be an obstacle in practice.  As demonstrated in \citet{angelica}, it is possible to derive a principal component basis (and the corresponding eigenvalues) empirically.

\section{The difficulty in measuring physical parameters}
\label{physicalparamsbad}
In the previous section, we saw that the foregrounds can be accurately described using only a small number of foreground eigenmodes.  Put another way, there are very few distinguishing features in radio foreground spectra, so there are really only a few independent parameters in the data.  In this section, we will see that this places severe constraints on our ability to measure the physical parameters listed in Table \ref{table1}.

We employ a Fisher matrix formalism to estimate the best possible error bars on measurements of these parameters.  To do so, we imagine that one has already used Equation \ref{leastsq} to compute an estimator $\mathbf{\hat{a}}$ of the expansion coefficient vector from the data\footnote{Instead of considering the expansion coefficients, one could instead deal with the foreground spectrum $\mathbf{x}$ and its estimator $\mathbf{\hat{x}}$ directly.  However, unlike for the expansion coefficients, the spectra themselves have the unfortunate property of having non-diagonal error covariances (\emph{e.g.} Equation \ref{wienererror} for Wiener filtering), which makes the subsequent analysis and interpretation in this section more cumbersome.}.  The precise value of this estimator will vary because of the random nature of instrumental noise, whose effect can be quantified by computing the error covariance matrix of $\mathbf{\hat{a}}$.  Since both the Wiener filtering method and the truncated least-squares method require as a first step a full least-squares estimation of $\mathbf{a}$, we can use the standard formula for the error covariance $\Sigma$ for least-squares fitting \citep{mapmaking}:
\begin{equation}
\Sigma \equiv \langle [\mathbf{\hat{a}}-\mathbf{\langle a\rangle}]  [\mathbf{\hat{a}}-\langle \mathbf{a} \rangle]^{\mathbf{t}} \rangle = \mathbf{[V^t N^{-1} V]^{-1}} = \kappa^2 \mathbf{I},
\end{equation}
where like before we used the fact that in our whitened units, $\mathbf{N} = \kappa^2 \mathbf{I}$.  Once again, the angle brackets $\langle \dots \rangle$ denote an ensemble average, or (equivalently from the standpoint of practical observations) an average over many independent lines-of-sight.  With this covariance, the probability distribution $L$ of $\mathbf{\hat{a}}$ is given by
\begin{equation}
\label{likelihoodfct}
L(\mathbf{\hat{a}}; \mathbf{\Theta}) = \frac{1}{\sqrt{(2\pi)^{N_c} \det \mathbf{\Sigma}}} \exp \left[ -\frac{1}{2} [\mathbf{\hat{a}}-\langle \mathbf{a} \rangle]^t \mathbf{\Sigma}^{-1} [\mathbf{\hat{a}}-\langle \mathbf{a}\rangle]  \right],
\end{equation}
where $\mathbf{\Sigma}$ is the error covariance matrix, and $\mathbf{\Theta} \equiv (\theta_1, \theta_2, \theta_3, \dots )$ is a vector of model parameters (such as the parameters in Table \ref{table1}), which enters our expression because the true expansion coefficient vector $\mathbf{a}$ depends on these parameters.  Interpreted as a function of $\mathbf{\Theta}$ for a given measurement $\mathbf{\hat{a}}$, $L$ is the so-called likelihood function for the parameters that we wish to constrain.  The tightness of our constraint is closely related to the Fisher information matrix, which is defined as
\begin{equation}
\label{fisherdef}
\mathbf{F}_{AB} \equiv \Bigg{\langle} \frac{\partial^2 \mathcal{L}}{\partial \theta_A \partial \theta_B} \Bigg{\rangle},
\end{equation}
where $\mathcal{L} \equiv - \ln L$.  If our estimator for the parameters is unbiased, \emph{i.e.}
\begin{equation}
\langle \mathbf{\Theta} \rangle = \mathbf{\Theta}_0,
\end{equation}
where $\mathbf{\Theta}_0$ is the true parameter vector, then the Cramer-Rao inequality states that the error bars on $\theta_A$ (defined by the standard deviation $\Delta \theta_A \equiv \sqrt{\langle \theta_A^2 \rangle - \langle \theta_A \rangle^2}$) satisfy
\begin{equation}
\Delta \theta_A \geq (\mathbf{F}^{-1})_{AA}^{1/2}
\end{equation}
if we estimate all the parameters jointly from the data.  Computing the Fisher matrix thus allows us to estimate the ability of an experiment to constrain physical parameters, with the covariance between the parameter estimates equal to $\mathbf{F}^{-1}$ if the data analysis is done in an optimal fashion.

\begin{figure}
\centering
\includegraphics[width=0.45\textwidth]{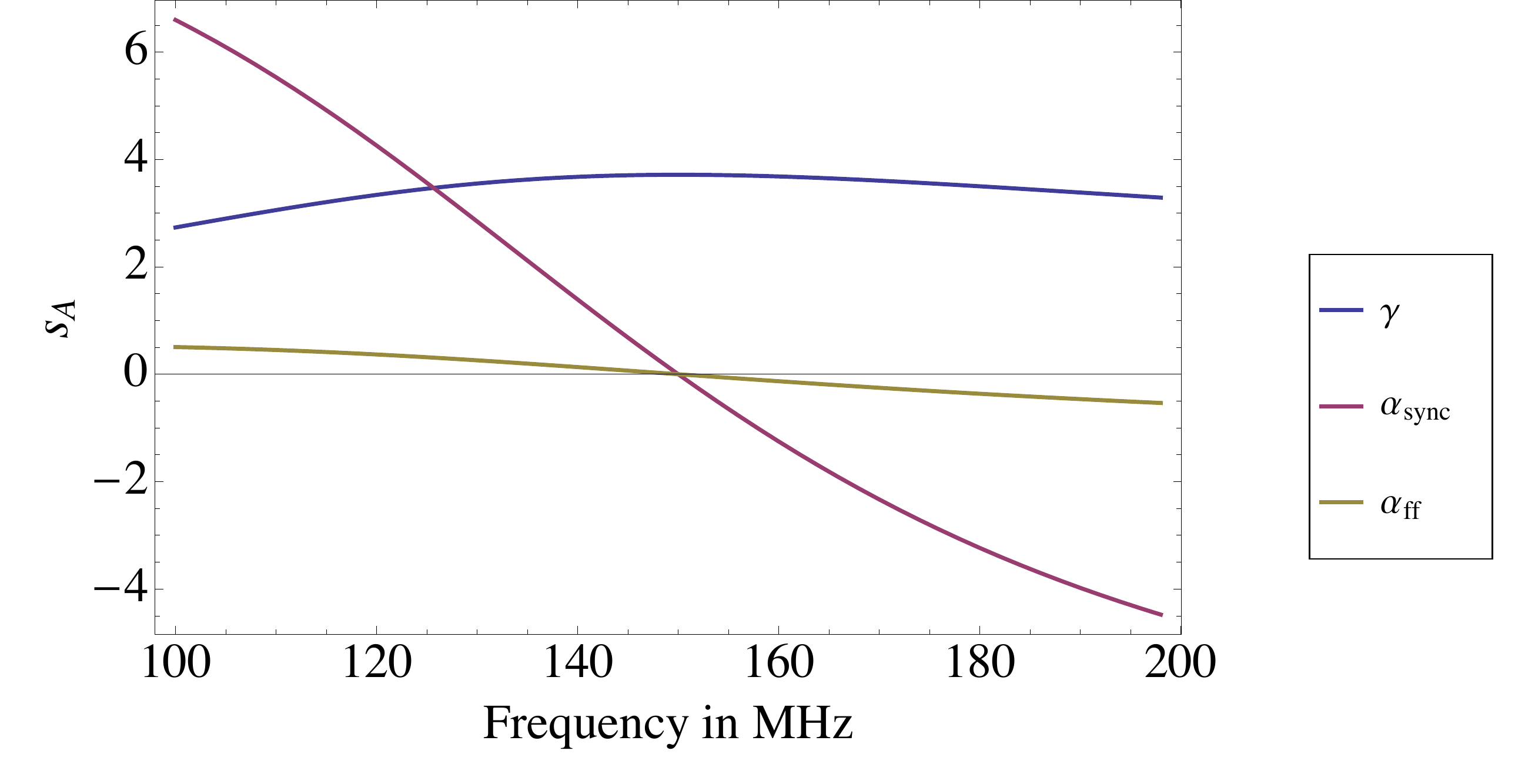}
\caption{Parameter derivative vectors $\mathbf{s}_A$ (Equation \ref{vectderiv}) for $\gamma$, $\alpha_{sync}$, and $\Delta \alpha_{sync}$.}
\label{paramderiv1}
\end{figure}
\begin{figure}
\centering
\includegraphics[width=0.45\textwidth]{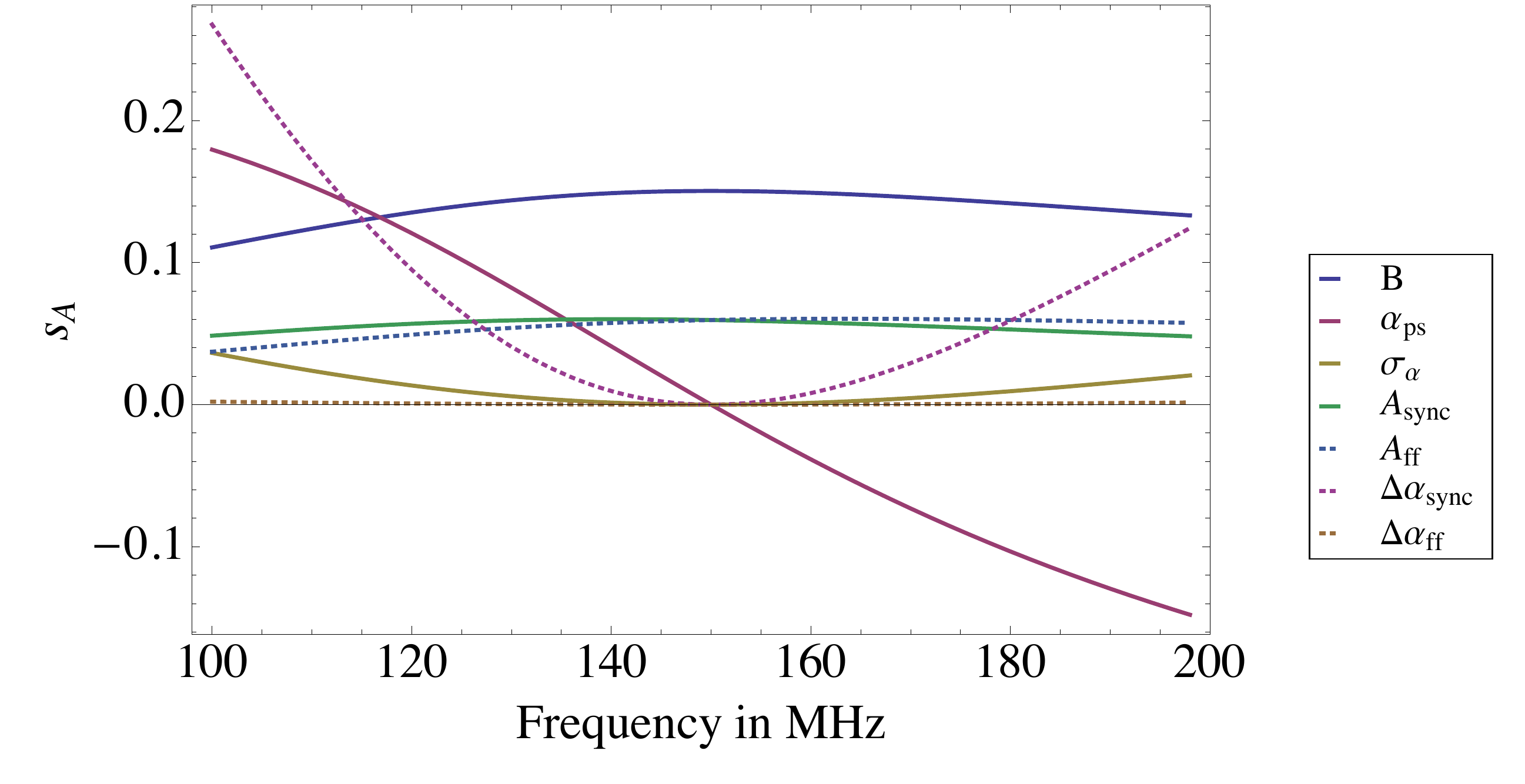}
\caption{Parameter derivative vectors $\mathbf{s}_A$ (Equation \ref{vectderiv}) for $B$, $\alpha_{ps}$, $\sigma_\alpha$, $A_{sync}$,  $A_{f\!f}$, $\alpha_{f\!f}$, and $\Delta \alpha_{f\!f}$.}
\label{paramderiv2}
\end{figure}

Let us now compute the Fisher matrix for the foreground parameters listed in Table \ref{table1}.  Inserting  Equation \ref{likelihoodfct} into Equation \ref{fisherdef} and performing some matrix algebra, one can show \citep{KLmodes} that the Fisher matrix reduces to
\begin{equation}
\mathbf{F}_{AB} = \frac{1}{2} \textrm{Tr} \left[ \mathbf{\Sigma}^{-1} \frac{\partial \mathbf{\Sigma}}{\partial \theta_A} \mathbf{\Sigma}^{-1} \frac{\partial \mathbf{\Sigma}}{\partial \theta_B}\right] + \frac{\partial \mathbf{\langle a^t \rangle}}{\partial \theta_A} \mathbf{\Sigma}^{-1} \frac{\partial \mathbf{\langle a\rangle}}{\partial \theta_B}.
\end{equation}
Referring back to our expression for $\mathbf{\Sigma}$, we see that the first term vanishes because $\mathbf{\Sigma}$ depends only on the noise level and not on the physical parameters.  This gives
\begin{equation}
\label{fish1}
\mathbf{F}_{AB} =\frac{1}{\kappa^2} \left( \frac{\partial \mathbf{\langle a^t \rangle}}{\partial \theta_A} \right)\!\!\left( \frac{\partial \mathbf{\langle a\rangle}}{\partial \theta_B}\right).
\end{equation}
Note that this expression depends on the mean vector $\mathbf{\langle a\rangle}$, not the estimator $\mathbf{\hat{a}}$.  This is because the Fisher matrix formalism tells us what the error bars are for the \emph{optimal} method, whatever that method happens to be.  We thus do not expect $\mathbf{\hat{a}}$ to appear in $\mathbf{F}_{AB}$, for if it did we would have the freedom to plug in a possibly sub-optimal estimator of our choosing, and by construction the Fisher matrix formalism contains information about the optimal errors.

With $\mathbf{\langle a \rangle}$ signifying the true expansion coefficient vector, we know from Equations \ref{defmean} and \ref{measurementeqn} that
\begin{equation}
\mathbf{\langle a \rangle} = \mathbf{V^t} \langle \mathbf{x} \rangle = \mathbf{V^t}\mathbf{m}.
\end{equation}
Inserting this into our expression for the Fisher matrix, we obtain\footnote{We have implicitly assumed that the whitening procedure performed in Equations \ref{nondim1} was with respect to some \emph{fiducial} foreground model, so that $\mathbf{V}$ does not depend on the foreground parameter vector $\mathbf{\Theta}$.}
\begin{equation}
\label{geomfisher}
\mathbf{F}_{AB} =\frac{1}{\kappa^2} \left( \frac{\partial \mathbf{m^t}}{\partial \theta_A} \right) \mathbf{V} \mathbf{V^t} \left( \frac{\partial \mathbf{m}}{\partial \theta_B}\right) = \frac{1}{\kappa^2} \mathbf{s}_A \cdot \mathbf{s}_B,
\end{equation}
where we have used the fact that $\mathbf{V^t} = \mathbf{V}^{-1}$ since the eigenforegrounds are orthogonal, and have defined
\begin{equation}
\label{vectderiv}
\mathbf{s}_A \equiv \frac{\partial \mathbf{m}}{\partial \theta_A}.
\end{equation}
Each component of Equation \ref{geomfisher} can be geometrically interpreted as a dot product between two $\mathbf{s}$ vectors in an $N_c$-dimensional space.  As $\mathbf{m}$ encodes the whitened foreground spectrum (Equation \ref{nondim1}), each $\mathbf{s}$ vector quantifies the change in the expected foreground spectrum with respect to a physical parameter.  And since the final covariance matrix on the parameter errors is given by $\mathbf{F}^{-1}$, the greater the dot product between two different $\mathbf{s}$ vectors, the more difficult it is to measure the two corresponding physical parameters.  The forms of different $\mathbf{s}_A$'s thus give intuition for the degeneracies in a set of parameters.

\begin{table*}
\caption{Dimensionless dot products between $\mathbf{s}_A$ vectors (Equation \ref{vectderiv}) for the foreground parameters listed in Table \ref{table1}, or equivalently, a normalized version of the Fisher matrix (given by Equation \ref{geomfisher}) so that the diagonal elements are unity.  The derivatives were evaluated at the fiducial foreground parameters for the foreground model described in Section \ref{fgmodel}.  The frequency range of the experiment was taken to go from $100\,\textrm{MHz}$ to $200\,\textrm{MHz}$, with $50$ equally spaced channels.  The matrix is symmetric by construction, so the bottom left half has been omitted for clarity.}
\begin{tabular}{p{1.0cm}|p{1.0cm}|p{1.0cm}|p{1.0cm}|p{1.0cm}|p{1.1cm}|p{1.0cm}|p{1.1cm}|p{1.0cm}|p{1.0cm}|p{1.0cm}}
 & B & $\gamma $& $A_{sync}$ &$ A_{f\!f}$ & $\alpha_{ps}$& $\alpha_{sync}$  &$\alpha_{f\!f}$  &$\sigma_{\alpha} $ & $\Delta \alpha_{sync}$ & $\Delta \alpha_{f\!f}$\\
 \hline
B & 1.00 & 1.00 & 0.998 & 0.998 & 0.0603 & 0.111 & -0.00425 & 0.683 & 0.664 & 0.705 \\
$\gamma $ & --- & 1.00 & 0.999 & 0.998 & 0.0603 & 0.111 & -0.00425 & 0.683 & 0.664 & 0.705 \\
$A_{sync}$ & --- & ---& 1.00 & 0.993 & 0.113 & 0.163 & 0.0489 & 0.696 & 0.680 & 0.713  \\
$A_{f\!f}$ & --- & ---& --- &1.00 &  -0.00413& 0.0465& -0.0684& 0.661& 0.639& 0.688 \\
$\alpha_{ps}$ & ---& ---& --- &--- & 1.00 & 0.997& 0.996& 0.333& 0.389& 0.254\\
$\alpha_{sync}$ & ---& ---& --- &--- & --- & 1.00 & 0.987& 0.399& 0.454& 0.322 \\
$\alpha_{f\!f}$  & --- & --- & --- &--- & --- & --- & 1.00 & 0.245& 0.303& 0.164\\
$\sigma_{\alpha}$  & ---& ---& --- &--- & ---& ---& --- & 1.00 & 0.998& 0.996 \\
$\Delta \alpha_{sync}$ & ---& ---& --- &--- & ---& ---& --- &---& 1.00 & 0.987\\
$\Delta \alpha_{f\!f}$ & ---& ---& --- &--- & ---& ---& --- &---& --- & 1.00 \\
\end{tabular}
\label{table2}
\end{table*}

\begin{table*}
\caption{Eigenvectors (Equation \ref{normfour}) of the normalized Fisher matrix (Table \ref{table2}).  Each row represents an eigenvector, and going from top to bottom the eigenvectors are arranged in descending value of eigenvalue.}
\begin{tabular}{p{1.0cm}|p{1.0cm}|p{1.0cm}|p{1.0cm}|p{1.0cm}|p{1.0cm}|p{1.0cm}|p{1.0cm}|p{1.0cm}|p{1.0cm}|p{1.0cm}}
 & B & $\gamma $& $A_{sync}$ &$ A_{f\!f}$ & $\alpha_{ps}$& $\alpha_{sync}$  &$\alpha_{f\!f}$  &$\sigma_{\alpha} $ & $\Delta \alpha_{sync}$ & $\Delta \alpha_{f\!f}$\\
 \hline
$\theta_1^\prime$ & 0.368& 0.368& 0.374& 0.36& 0.14& 0.164& 0.108& 0.366& 0.365& 0.366 \\
$\theta_2^\prime$ & 0.184& 0.184& 0.155& 0.219& -0.532& -0.522& -0.542& -0.053& -0.088& -0.005 \\
$\theta_3^\prime$ & 0.283& 0.283& 0.293& 0.275& 0.167& 0.13& 0.215& -0.442& -0.429& -0.455 \\
$\theta_4^\prime$ & 0.006& 0.006& 0.002& 0.01& -0.007& -0.046& -0.049& 0.089& 0.665& -0.739 \\
$\theta_5^\prime$ & -0.001& -0.001& 0.001& 0.& 0.002& 0.088& -0.087& -0.81& 0.472& 0.327 \\
$\theta_6^\prime$ &-0.001& -0.001& -0.018& 0.019& 0.81& -0.293& -0.506& 0.004& -0.041& 0.008 \\
$\theta_7^\prime$  & 0.247& 0.248& -0.125& -0.369& -0.097& 0.646& -0.534& 0.05& -0.081& -0.072\\
$\theta_8^\prime$  & 0.408& 0.41& -0.236& -0.58& 0.057& -0.411& 0.311& -0.033& 0.049& 0.043 \\
$\theta_9^\prime$ & 0.181& 0.114& -0.823& 0.524& 0.004& 0.048& 0.026& 0.& -0.003& 0.\\
$\theta_{10}^\prime$ & 0.7& -0.712& 0.038& -0.026& 0.& -0.002& -0.001& 0.& 0.& 0. \\
\end{tabular}
\label{table3}
\end{table*}

Shown in Figures \ref{paramderiv1} and \ref{paramderiv2} are plots of the $\mathbf{s}_A$ vectors for the foreground parameters shown in Table \ref{table1}.  For clarity, we have separated the derivatives into two plots that have different vertical scales.  Many of the parameters derivatives have similar shapes, suggesting that they will have a large ``dot product" in Equation \ref{geomfisher} and therefore large degeneracies between them.  Note that the overall normalization of the curves is irrelevant as far as degeneracies are concerned.  This is because two parameters with identically shaped curves but different normalizations are still completely degenerate as one can perfectly compensate for changes in one parameter with smaller (or larger) changes in the other.  Thus, to quantify the degree of degeneracy, we can form a matrix of normalized dot products between the $\mathbf{s}_A$ vectors, where the magnitude of each vector is individually normalized to unity.  The results are shown in Table \ref{table2}, where we see that the parameters form three degenerate groups --- the normalization parameters ($B$, $\gamma$, $A_{sync}$, $A_{f\!f}$), forming a degenerate $4\times 4$ block of $\approx$1's in the top left corner; the spectral index parameters ($\alpha_{ps}$, $\alpha_{sync}$,  $\alpha_{f\!f}$), forming a degenerate block in the middle; and the frequency coherence parameters ($\sigma_\alpha$, $\Delta \alpha_{sync}$, $\Delta \alpha_{f\!f}$), forming a degenerate block in the bottom right corner.  Indeed, computing the eigenvalues of the normalized Fisher matrix (\emph{i.e.} of Table \ref{table2}), one finds only three eigenvalues of order unity and higher, with the next largest eigenvalue of order $10^{-3}$.  This suggests that there are really only three independent foreground parameters that can be measured.

To gain an intuition for what these independent parameters quantify, consider the first few eigenvectors of the normalized Fisher matrix $\mathbf{\widehat{F}}$ shown in Table \ref{table2}, that is, vectors that satisfy
\begin{equation}
\label{normfour}
\mathbf{\widehat{F}} \mathbf{\theta_n^\prime} = \lambda_n \mathbf{\theta_n^\prime}.
\end{equation}
These parameter eigenvectors are listed in Table \ref{table3}, where each row gives the weighted average of the original parameters that one must take to form the ``eigenparameters".  To see what foreground parameters are well characterized by $21\,\textrm{cm}$ tomography, we look at the first few eigenvectors, which can be measured with the highest signal-to-noise ratio:
\begin{eqnarray}
\theta_1^\prime &\approx & 0.4 (B+ \gamma  +\sigma_\alpha + A_{sync} + A_{f\!f} +\Delta \alpha_{sync} + \Delta \alpha_{f\!f} ) \nonumber \\
&&+ 0.1 (\alpha_{ps}+\alpha_{f\!f}) +0.2 \alpha_{sync}\nonumber \\
\theta_2^\prime &\approx & 0.2( B+\gamma+A_{f\!f}+ A_{sync}) - 0.1(\sigma_\alpha + \Delta \alpha_{sync})  \nonumber \\
&& -0.5 (\alpha_{ps} + \alpha_{sync} + \alpha_{f\!f})-0.02 \Delta \alpha_{f\!f}  \nonumber \\
\theta_3^\prime &\approx & 0.3 (B +\gamma + A_{sync} + A_{f\!f}) +0.2 (\alpha_{ps} + \alpha_{f\!f}) + 0.1 \alpha_{sync} \nonumber \\
&&-0.4 (\sigma_\alpha + \Delta \alpha_{sync})-0.5\Delta \alpha_{ff} ,
\end{eqnarray}
where for clarity the coefficients have been rounded to one significant figure (see Table \ref{table3} for more precise values).  The first eigenparameter weights all the parameters equally except for the spectral indices, which are somewhat downweighted.  The first eigenparameter is thus roughly an ``everything but spectral indices" parameter.  Crudely speaking, this parameter is a generalized normalization parameter because it is mostly comprised of normalization parameters ($B$, $\gamma$, $A_{sync}$, $A_{f\!f}$) and frequency coherence parameters, whose spectral effects are not apparent until the frequency range is large.  Examining the third eigenvector, we see that it is most heavily weighted towards the frequency coherence parameters, suggesting that we do have an independent parameter that acts as a ``generalized frequency coherence".  However, being the third eigenvector, our ability to make measurements of it is lower than with the first eigenvector or the second eigenvector, which is a ``generalized spectral index".

To see which degrees of freedom we cannot constrain in our foreground model, we consider the last couple of eigenparameters:
\begin{eqnarray}
\theta_9^\prime &\approx & 0.18 B + 0.11\gamma-0.82A_{sync}+0.52A_{f\!f}\nonumber \\
\theta_{10}^\prime &\approx & 0.7 B - 0.71 \gamma,
\end{eqnarray}
where we have omitted all terms with coefficients less than $0.1$.  The last eigenvector is dominated by $B$ and $\gamma$, which appear with almost equal and opposite magnitudes.  This suggests that differences between $B$ and $\gamma$ are extremely hard to measure.  In the penultimate eigenvector, the sum of coefficients for $B$, $\gamma$, and $A_{f\!f}$ is roughly equal and opposite to the coefficient of $A_{sync}$.  This implies that the collective difference between $A_{sync}$ and the linear combination of $B$, $\gamma$, and $A_{f\!f}$ is also difficult to tease out from the data, but less so than the difference between $B$ and $\gamma$.

That we can only measure a small number of foreground parameters independently is perhaps unsurprising, given that in Figure \ref{eigenvalueswnonoise} we saw that most of the foreground power comes from a small number of eigenforeground modes that lack distinctive features.  Note that since $\kappa$ enters our expression for the foreground parameter Fisher matrix (Equation \ref{geomfisher}) as an overall multiplicative constant, instrumental noise has no effect on the foreground parameter degeneracies we examined in this section.  The parameter degeneracies are due purely to the form of the foregrounds, and have nothing to do with the noise.  On the other hand, when quantifying the effective number of parameters in the Wiener filter (Equation \ref{neff}) or when solving for the optimal number of eigenforeground modes to measure in the truncated least-squares method (Equation \ref{optm}), our expressions explicitly depend on the noise level.  For instance, in the limit of a noiseless experiment the best characterization of foregrounds is obtained by measuring all the eigenmodes, not the three or four modes found in Sections \ref{wiener} and \ref{TLS}.  In general, however, the results of this section show that one must be able to measure at least three eigenmodes at a reasonable signal-to-noise in order to adequately model the foreground spectrum.  If Equation \ref{neff} or Equation \ref{optm} suggest that fewer eigenmodes should be used in the model, it is simply because one's experiment is too noisy to allow the foregrounds to be measured accurately.

\section{Conclusions}
\label{conc}
In this paper, we have shown that despite the complicated dependence that low-frequency radio foregrounds may have on physical parameters, the resulting spectra are always rather generic and featureless.  The bad news from this came in Section \ref{physicalparamsbad}, where we saw that even the most careful foreground spectrum measurements are unlikely to yield interesting constraints on foreground physics, thanks to high levels of degeneracies between different foreground parameters.  This, however, is good news for those who simply consider foregrounds an impediment to $21\,\textrm{cm}$ tomography.  In Section \ref{fewparamsfine} we saw that it was precisely because the foreground spectra were so featureless that they could be characterized to a greater accuracy than necessary for foreground subtraction using only three or four independent parameters.  This bodes well for $21\,\textrm{cm}$ astrophysics and cosmology, for it suggests that extremely detailed and physically motivated foreground models will not be necessary for successful foreground cleaning.
\section*{Acknowledgments}
The authors would like to thank Joshua Dillon, Leo Stein, and Christopher Williams for useful discussions.  This work was supported by NSF grants AST-0907969, AST-0708534.

\bibliographystyle{mn2e}
\bibliography{foregroundparam}

\end{document}